\documentclass[twocolumn]{aastex631}
\usepackage{newunicodechar}
\newunicodechar{ }{\,}
\usepackage{amsmath} 

\begin{document}

\title{Warped Disk Galaxies: \\Statistical Properties from DESI Legacy Imaging Surveys DR8}

\author[0009-0004-6972-0193]{Yiheng Wang}
\affiliation{Purple Mountain Observatory, Chinese Academy of Sciences, Nanjing 210023, Jiangsu, China}
\affiliation{School of Astronomy and Space Sciences, University of Science and Technology of China, Hefei, Anhui 230026, China}

\author[0009-0000-8850-0250]{Han Qu}
\affiliation{Purple Mountain Observatory, Chinese Academy of Sciences, Nanjing 210023, Jiangsu, China}
\affiliation{School of Astronomy and Space Sciences, University of Science and Technology of China, Hefei, Anhui 230026, China}

\author[0000-0002-8817-4587]{Jiafeng Lu}
\affiliation{Institute for Astronomy, School of Physics, Zhejiang University, Hangzhou 310058, Zhejiang, China}
\affiliation{Center for Cosmology and Computational Astrophysics, Zhejiang University, Hangzhou 310058, Zhejiang, China}

\author[0000-0002-4911-6990]{Huiyuan Wang}
\affiliation{Department of Astronomy, University of Science and Technology of China, Hefei, 230026, China}
\affiliation{School of Astronomy and Space Sciences, University of Science and Technology of China, Hefei, Anhui 230026, China}

\author[0000-0003-1588-9394]{Enci Wang}
\affiliation{Department of Astronomy, University of Science and Technology of China, Hefei, 230026, China}
\affiliation{School of Astronomy and Space Sciences, University of Science and Technology of China, Hefei, Anhui 230026, China}

\author[0000-0002-5458-4254]{Xi Kang}
\affiliation{Institute for Astronomy, School of Physics, Zhejiang University, Hangzhou 310058, Zhejiang, China}
\affiliation{Center for Cosmology and Computational Astrophysics, Zhejiang University, Hangzhou 310058, Zhejiang, China}
\affiliation{Purple Mountain Observatory, Chinese Academy of Sciences, Nanjing 210023, Jiangsu, China}

\correspondingauthor{Xi Kang}
\email{kangxi@zju.edu.cn}

\begin{abstract}

Warped structures are often observed in disk galaxies, yet their physical origin is still under investigation. We present a systematic study of warped edge-on disk galaxies based on imaging data from the DESI Legacy Imaging Surveys DR8, with the expectation that this large sample size, enabled by wide-area surveys, will offer new perspectives on the formation of disk warps. Using a deep learning approach, we trained an EfficientNet-B3 convolutional neural network to classify the morphology of edge-on-disk galaxies into warped and non-warped categories. Our model was trained on a curated and visually verified set of labeled galaxy images and applied to a large dataset of over 595,651 edge-on disk galaxies selected from the Galaxy Zoo DESI catalog. Our results provide the largest warp catalog to date, consisting of 23996 warped edge-on disk galaxies, and reveal statistical trends between warp occurrence and galaxy properties. Compared to their non-warped counterparts, these warped disk galaxies tend to have bluer colors, lower stellar masses, higher gas fractions and star-formation rates, smaller Sérsic indices and larger disk sizes. In addition, warped disk galaxies show higher projected number densities of neighboring galaxies than their non-warped counterparts, particularly within \( R_{\mathrm{proj}} \lesssim 50~\mathrm{kpc} \), where the local number density is roughly twice as high.

\end{abstract}




\section{Introduction}\label{sec:intro}

Galaxies display diverse structures in their morphology, including spiral arms, stellar bars, central bulges, and tidal substructures. These morphological features serve as both tracers and drivers of galaxy evolution. Bulges influence galaxy evolution by stabilizing the central regions, regulating gas inflows, and often correlating with star formation quenching and black hole growth \citep[e.g.,][]{kormendy_2004, Bluck_2014, Kormendy_2013}, while bars drive gas from the disk toward the center \citep{sakamoto_1999}, fueling star formation and potentially triggering active galactic nuclei \citep{shlosman_1989, sheth_2005}, thus playing a crucial role in secular evolutionary processes. Tidal substructures may indicate past interactions that dynamically reshape galaxy structure \citep{volker_1999, johnston_1999, Ibata_2001}. Since these processes directly link morphology to galactic evolutionary histories, studying galaxy morphology becomes crucial for uncovering the physical mechanisms behind their formation and evolution.

Among various morphological features, structural asymmetries provide particularly valuable insights into the dynamical and evolutionary states of galaxies. Such asymmetries can arise from internal instabilities or external perturbations, revealing ongoing processes that shape disk evolution. A notable manifestation of these asymmetries is the presence of warped disks, which are characterized by systematic vertical displacements of the stellar and gaseous components from an ideal flat plane, typically with amplitudes of a few degrees. This phenomenon has been observed in our Galaxy \citep[e.g.,][]{burke_1957, jonsson_2024, kerr_1957, Djorgovski_1989, Nurhidayat_2020, Hunt_2025} and in many galaxies in the local Universe \citep[e.g.,][]{Reshetnikov_1995, Reshetnikov_1998, Garcia_2002, Saroon_2022}. \citet{briggs_1990} provided a comprehensive summary and classification of the observed characteristics of galactic warps. Moreover, recent observations suggest that warp frequencies increase with redshift \citep{Reshetnikov_2025}, implying an evolving nature of disk asymmetries over cosmic time. 

These observational results highlight that warps are a common and long-lived feature in disk galaxies, yet the physical mechanisms responsible for their origin and longevity remain to be fully understood. Proposed explanations include internal bending instabilities\citep{Sellwood_1996, Binney_1992}, misaligned dark matter halos\citep{Sparke_1998, Debattista_1999, han_2023a, han_2023b}, satellite interactions\citep{Weinberg_1998, Laporte_2018, binney_2024}, and ongoing cosmic gas accretion\citep{Rok_2010, Voort_2015}. Several studies have also highlighted a possible link between environmental density and the presence of warps, with more strongly warped galaxies often found in denser regions\citep{Reshetnikov_1999}.

The unclear origin of galactic warps is partly due to the small and non-uniform samples available, which limit reliable statistical analyses of their connection to galaxy structure and environment\citep[e.g.,][]{Garcia_2002, Sanchez_2003, Ann_2006, Skryabina_2024, Reshetnikov_2025}. To bridge this gap, we carry out a comprehensive analysis based on the DESI Legacy Imaging Surveys (DR8). The wide sky coverage, imaging depth, and relatively high resolution of the survey make it possible to identify a large number of warped galaxy candidates. Using a machine learning classifier applied to edge-on disk galaxies from the Galaxy Zoo DESI catalog, we identify 23,996 highly-confidence warped and 288,562 non-warped galaxies out of a total of 595,651 edge-on disk galaxies. We find statistical differences across several galaxy properties, including stellar mass, gas fraction, color, star formation rate, and disk morphology such as bulge prominence and disk size. These differences suggest that the presence of warps is linked to systematic variations in the structural and stellar characteristics of disk galaxies.

This paper is organized as follows. Section~\ref{sec:meth} introduces the dataset, explains how the sample was selected, and describes the training process of the machine learning model. In Section ~\ref{sec:result} we analyze the systematic differences in various properties between warped and non-warped edge-on-disk galaxies predicted by our classifier. Finally, in Section~\ref{sec:diss}, we summarize our findings and discuss their implications for understanding the nature of galactic warps. We note that both the training and prediction processes are based solely on optical imaging, without incorporating morphological features of the gaseous disk.

\section{Methodology}\label{sec:meth}

\subsection{Surveys and Data}\label{subsec:data}

The Dark Energy Spectroscopic Instrument Legacy Imaging Surveys (DESI-LS) provide high-quality, wide-field optical imaging, enabling the detection of faint structural features such as spiral arms, bars, tidal debris, and disk warps. It is composed of three individual surveys: the Dark Energy Camera Legacy Survey (DECaLS), the Beijing-Arizona Sky Survey (BASS), and the Mayall z-band Legacy Survey (MzLS). BASS and MzLS (jointly referred as BASS/MzLS) together cover the Northern Galactic Cap (NGC) from Kitt Peak, USA; using the Bok 2.3-meter and Mayall 4-meter telescopes. DECaLS uses the 4-meter Blanco telescope at Cerro Tololo Inter-American Observatory equipped with the Dark Energy Camera (DECam; \cite{flaugher_dark_2015}) to image the Southern Galactic Cap (SGC) and the $\delta \leq 34^\circ$ region of the NGC. The combination of DECaLS and BASS/MzLS provides 14,000 $deg^2$ grz imaging data for DESI targeting. Notably, the Dark Energy Survey (DES) is conducted using the same instrument as DeCaLS, so its imaging data are also included in the DESI Legacy Surveys Data Release 8 (DR8). Together, these four surveys, DECaLS, BASS, MzLS, and DES, provide a combined sky coverage of 19,437 $deg^2$. 

To ensure uniformity across the entire footprint, the imaging properties of these four surveys are similar by design, including depth and quality of the image across all components\citep{dey_overview_2019}. The median coadded depths (5$\sigma$ detection of a point source) reach approximately g = 24.2, r = 23.8, and z = 23.3 for BASS/MzLS, and g = 24.8, r = 24.2, and z = 23.4 for DECaLS. DES DR2 provides a median coadded catalog depth of g = 24.7, r = 24.4, and z = 23.1 for a 1$''$.95 diameter aperture in signal-to-noise ratio = 10.

The uniform, deep and wide imaging of these surveys enables large-scale morphology classification by training deep learning models on volunteer-labeled subsets and applying them to the remaining images. \citet{walmsley_2021} presented the Galaxy Zoo DECaLS project, the first to produce a large-scale catalog of morphological measurements for all Galaxy Zoo questions by training deep learning algorithms on citizen scientist classifications. They trained their model on DECaLS images using volunteer labels for various morphological features from the Galaxy Zoo website and then made the corresponding predictions. This catalog provided high-accuracy morphological classifications for over 314,000 galaxies from the DECaLS imaging surveys DR5 and within the SDSS DR8 footprint.

Building on the uniformly deep and consistent imaging dataset, \citet{walmsley_2023} extended the Galaxy Zoo morphological classification to the full DESI-LS footprint, expanding the sky coverage from about 5000 to over 19,000 deg\textsuperscript{2}, covering nearly the entire extragalactic sky and enabling synergy with other major surveys such as ALFALFA and MaNGA. The released Galaxy Zoo DESI catalog includes fainter ($r<19.0$ vs. $r<17.77$), smaller, and higher-redshift (up to $z\lesssim0.4$ vs. $z<0.15$) galaxies compared to Galaxy Zoo DECaLS. Galaxies were selected with $r<19.0$ (to ensure sufficient resolution for visual morphology), surface brightness $\mu > 18$ mag arcsec$^{-2}$, and non-PSF profiles (based on the Tractor catalog), yielding an initial sample of 8,956,477 sources. After discarding images with more than $20\%$ missing flux in any band, 8,689,370 galaxies remained as the final sample. For these galaxies, automated morphology predictions were released, providing the predicted fraction of volunteer responses for each Galaxy Zoo question, typically accurate to within $5\text{--}10\%$ for each possible answer.

\subsection{Sample Selection\label{subsec:sample_selection}}
\begin{table*}[htbp]
\centering
\footnotesize  
\caption{Predicted morphological parameters based on Galaxy Zoo DESI classifications. All fractions are in the range $[0,1]$.}
\begin{tabular}{l >{\raggedright\arraybackslash}p{12cm}}  
\hline
\hline
Parameter & Physical Meaning \\
\hline
\texttt{edge-on\_yes\_fraction} & Predicted fraction (0--1) of votes classifying the galaxy as edge-on. \\
\texttt{merging\_minor-disturbance\_fraction} & Predicted fraction (0--1) of votes indicating minor disturbances, likely due to a minor merger. \\
\texttt{merging-none\_fraction} & Predicted fraction (0--1) of votes indicating no signs of merging or disturbance. \\
\texttt{bulge-size\_small\_fraction} & Predicted fraction (0--1) of votes classifying the galaxy's bulge as small relative to total stellar mass or light. \\
\texttt{bulge-size\_moderate\_fraction} & Predicted fraction (0--1) of votes classifying the galaxy's bulge as moderate in size. \\
\hline
\end{tabular}
\label{tab:morph_params}
\end{table*}

Since our warp identification relies on optical imaging, we restrict our sample to edge-on galaxies, where the vertical displacement of the disk caused by warps is clearly observable. In less inclined galaxies, such distortions are largely hidden along the line of sight, making warp detection in optical images difficult. Following the recommended value in Table 2 of \citet{walmsley_2021}), we selected sources from the Galaxy Zoo DESI catalog by applying a threshold of \texttt{edge-on\_yes\_fraction} $\geq 0.8$. Applying this selection criteria, we obtained a sample 595,651 of edge-on disk candidates. Then we downloaded cutout images of each source using the DESI-LS cutout service. Cutout images were obtained at a fixed pixel scale of 0.262 arcsec/pixel, with view sizes tailored to each galaxy's angular size (see Appendix \ref{app:view}). To maintain consistent coverage across the dataset, our view size selection ensures that most edge-on galaxies fill a similar proportion of the cutout. This helps preserve the outer disk regions, even in galaxies with large bulges and concentrated central light.

We also utilize several other morphological fractions predicted from Galaxy Zoo DESI classifications to characterize the structural and merger-related properties of galaxies. These include, for example, \texttt{merging\_minor-disturbance\_fraction}, \texttt{merging-none\_fraction}, and bulge-size related fractions such as \texttt{bulge-size\_small\_fraction} and \texttt{bulge-size\_moderate\_fraction}. The definitions and physical meanings of all such parameters used in this work are summarized in Table~\ref{tab:morph_params}.

\subsection{Training Dataset Construction}

During the construction of the training and validation sets, one of the main challenges we encountered was the identification of a sufficient number of well-labeled warped galaxies, those exhibiting one-sided (U-shape) or two-sided (S-shape) warps, since such systems are relatively rare among edge-on galaxies when identified solely from optical-band images without accounting for gas-disk distributions. Moreover, warp features are also difficult to simulate or inject into real galaxy images without introducing unrealistic artifacts. For these reasons, we used visually identified real galaxies as positive warp examples to ensure morphological fidelity. 

\begin{figure*}[!htbp]
  \centering
  \includegraphics[width=\textwidth,height=0.95\textheight,keepaspectratio]{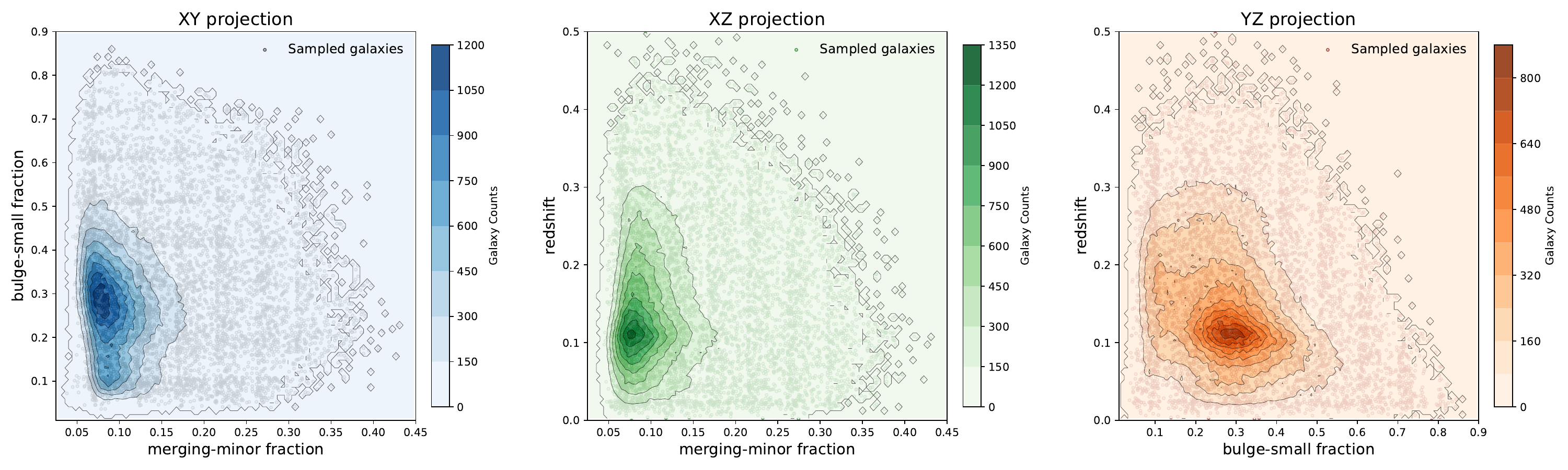}
  \caption{Three two-dimensional projections of edge-on galaxies in the parameter space of \texttt{merging\_minor-disturbance\_fraction}, \texttt{bulge-size\_small\_fraction}, and redshift. Contours indicate the distribution of all edge-on galaxies in each projected plane, and scatter points show the subset selected for labeling using our 3D binned sampling strategy, with up to 20 galaxies randomly drawn per bin. This approach ensures that rare types of galaxies are represented, improving the completeness and diversity of the training set.}
  \label{fig:3d_sample}
\end{figure*}

Initially, we suspected that the morphological signatures of disk warps might be partly interpreted by the Galaxy Zoo DESI model as minor disturbances or asymmetries. Therefore, we examined edge-on galaxies with relatively high \texttt{merging\_minor-disturbance\_fraction} values to search for potential warped systems that could have been classified as such by the machine predictions. We indeed found that warp-like features often exist among these galaxies, highlighting the potential of identifying a large number of warped systems in the optical bands. However, to ensure that the final training set was unbiased and representative of the full diversity of galaxy properties, we adopted a three-dimensional binned sampling strategy (see Figure~\ref{fig:3d_sample}). Specifically, we binned all edge-on galaxies in the parameter space of \texttt{merging\_minor-disturbance\_fraction}, \texttt{bulge-size\_small\_fraction}, and redshift, and then randomly selected 20 galaxies from each bin for labeling, or included all galaxies in bins containing fewer than 20 objects. This approach ensured that rare types of galaxies were adequately represented and prevented sampling bias toward more numerous systems. Labeling was performed by a single annotator, with ambiguous cases excluded, and clear criteria and high-resolution images were used to maintain consistency. 

\begin{figure}[!htbp]
  \centering
  \includegraphics[width=\columnwidth]{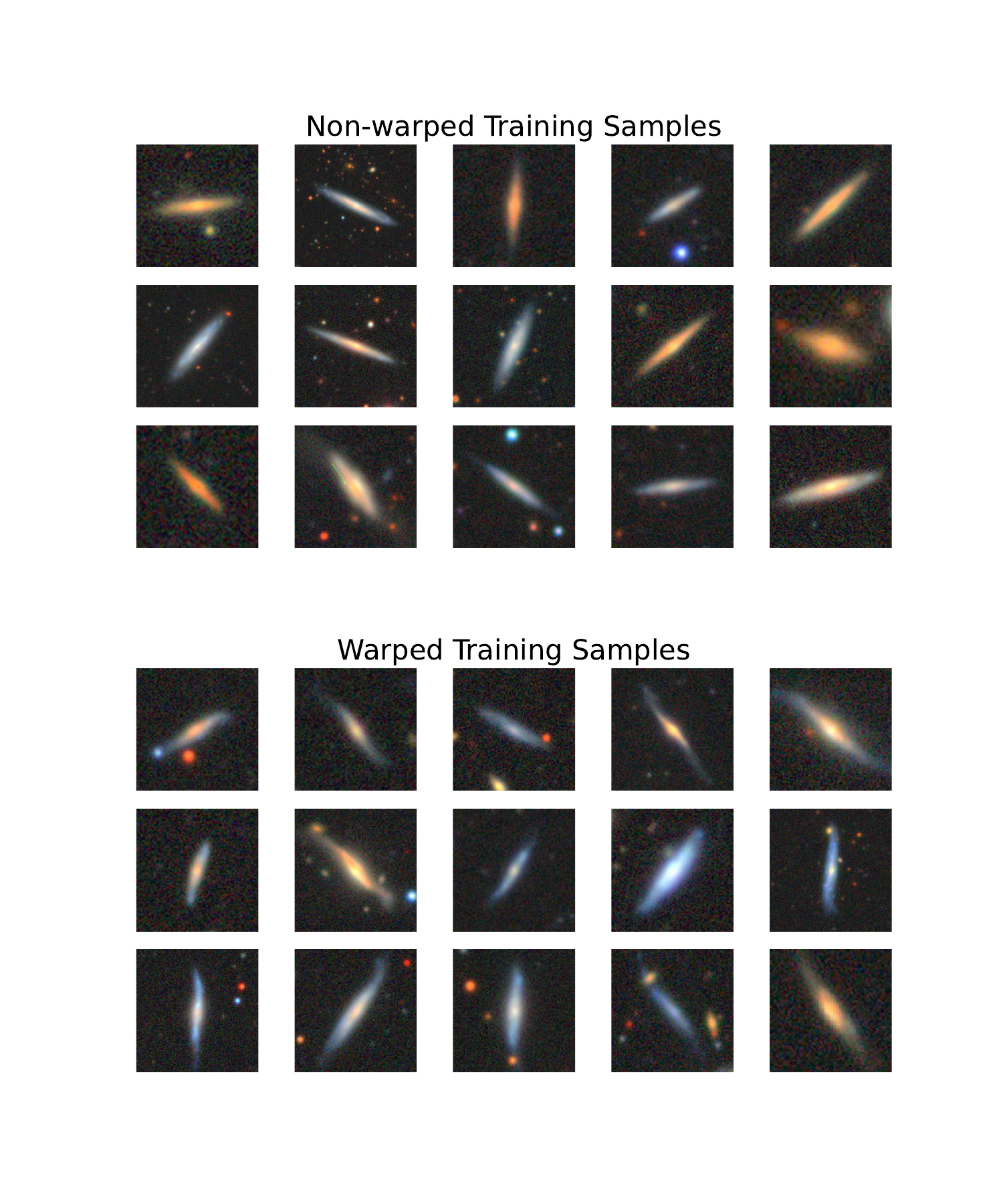}
  \caption{The figure shows example galaxy images used for training. The top panel shows non-warped galaxies, and the bottom panel shows warped galaxies.}
  \label{fig:training_sample}
\end{figure}

Finally, our labeled set consists of 2225 positive samples (S-shaped and U-shaped warped galaxies) and 3580 negative samples (non-warped edge-on galaxies). We randomly split the data into $80\%$, $10\%$, and $10\%$ for training, validation, and testing, respectively. Figure~\ref{fig:training_sample} illustrates a representative subset of the galaxies used for training the warp classification model, with non-warped examples shown in the top panel and warped examples in the bottom panel.

\subsection{Data Preprocessing and Model Training}
We adopt the EfficientNet-B3 architecture \citep{tan_19a} as the backbone of our classification model. EfficientNet employs a compound scaling method that uniformly adjusts network depth, width, and input resolution, achieving high accuracy with relatively low computational cost. The B3 variant provides an optimal balance between model capacity and efficiency, with approximately 12 million parameters and an input resolution of 300×300 pixels. We initialize the network with ImageNet-pretrained weights, enabling effective transfer learning by leveraging features learned from a large and diverse dataset. This allows the model to capture both low-level textures, such as local brightness gradients and subtle edge distortions, and higher-level structural features, including the global shape and orientation of galaxy disks. 

To increase diversity and reduce overfitting, we applied a series of data augmentation techniques to the galaxy images. The images were resized to 300×300 pixels and randomly flipped horizontally. We also applied random rotations up to 45 °, and color jittering with moderate adjustments in brightness, contrast saturation, and hue. Finally, the images were converted to tensors and normalized with a mean of 0.05 and a standard deviation of 0.1. The model training was optimized using Optuna\citep{optuna_2019}, which searched for the best hyperparameters within a predefined range to maximize performance. The search space included learning rate, dropout, class weight, and batch-size. We also experimented with a gradual unfreezing strategy—training only the classification head first and progressively unfreezing deeper layers—but found that it had little effect on the model performance. For the final model training, we used a batch size of 32 and trained for 100 epochs. Class weights were set to [1.0, 1.0], and a dropout rate of 0.297 was applied to prevent overfitting. The model was optimized using the AdamW optimizer with a learning rate of $1.7 \times 10^{-4}$ and a weight decay of $1.0 \times 10^{-2}$.

\begin{figure}
  \centering
  \includegraphics[width=\columnwidth]{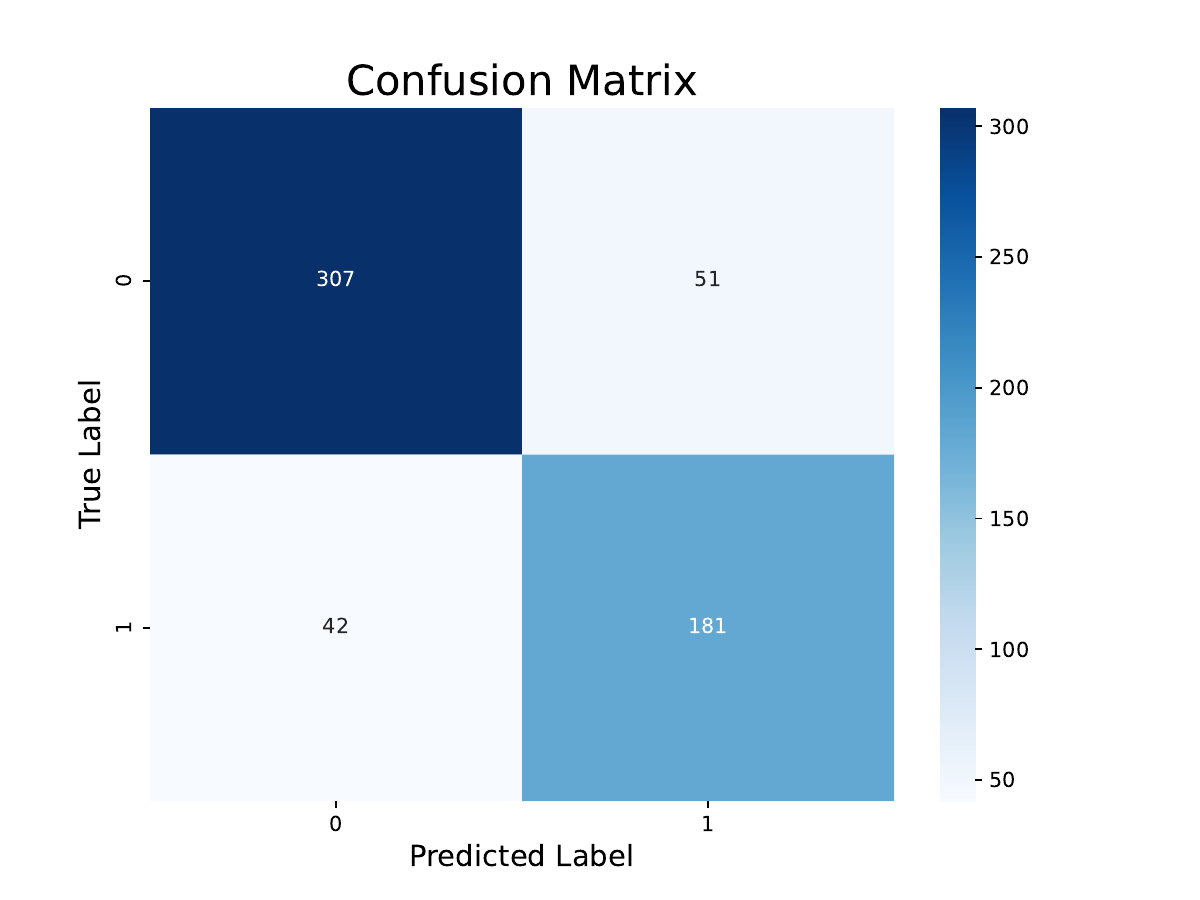}
  \caption{Confusion matrix of the best-performing model among all saved epochs.}
  \label{fig:cm}
\end{figure}

We saved models from all epochs and selected the one with best performance on the test set as the final model. The corresponding confusion matrix for the best model is shown in Figure~\ref{fig:cm}, illustrating how predictions are distributed across the two classes.

The evaluation metrics used in Table~\ref{tab:classification_report} are defined as follows. 
We denote one class as the positive class ($P$) and the other as the negative class ($N$). 
In our study, the positive class corresponds to warped galaxies. 
We define:

\begin{itemize}
    \item $TP$: number of positive samples correctly predicted as positive.
    \item $FP$: number of negative samples incorrectly predicted as positive.
    \item $FN$: number of positive samples incorrectly predicted as negative.
    \item $TN$: number of negative samples correctly predicted as negative.
\end{itemize}

Based on these, the metrics for the positive class are
\begin{align*}
\text{Precision}_P & = \frac{TP}{TP + FP}, \\
\text{Recall}_P    & = \frac{TP}{TP + FN}, \\
\text{F1-score}_P  & = \frac{2 \, \text{Precision}_P \, \text{Recall}_P}{\text{Precision}_P + \text{Recall}_P}, \\
\text{Support}_P   & = TP + FN.
\end{align*}

Similarly, the metrics for the negative class can be computed by swapping $P$ and $N$.

The overall accuracy is
\[
\text{Accuracy} = \frac{TP + TN}{TP + TN + FP + FN}.
\]

The macro-averaged and weighted-averaged metrics across the two classes are
\begin{align*}
\text{Macro Avg}    & = \frac{\text{metric}_P + \text{metric}_N}{2}, \\
\text{Weighted Avg} & = \frac{\text{Support}_P \cdot \text{metric}_P + \text{Support}_N \cdot \text{metric}_N}{\text{Support}_P + \text{Support}_N},
\end{align*}
where $\text{metric}_P$ and $\text{metric}_N$ can be Precision, Recall, or F1-score.

\begin{table}
\centering
\begin{tabular}{lcccc}
\hline
Class & Precision & Recall & F1-score & Support \\
\hline
0     & 0.88      & 0.86   & 0.87     & 358     \\
1     & 0.78      & 0.81   & 0.80     & 223     \\
\hline
Accuracy & \multicolumn{4}{c}{0.84} \\
Macro Avg & 0.83   & 0.83   & 0.83     & 581     \\
Weighted Avg & 0.84 & 0.84 & 0.84     & 581     \\
\hline
\end{tabular}
\caption{Precision, recall, F1-score, and support for each class on the test set. This corresponds to the best-performing model among all saved epochs.}
\label{tab:classification_report}
\end{table}

The model achieves an overall accuracy of 84\% on the test set (Table~\ref{tab:classification_report}). For class 0, the precision, recall, and F1-score are 0.88, 0.86, and 0.87, respectively, indicating a strong performance in identifying this class. For class 1, the precision recall and F1-score is 0.78, 0.81, and 0.80, showing slightly lower performance but still acceptable. The macro-averaged metrics are all 0.83, indicating balanced performance across classes. The weighted averages are similar to the overall accuracy.  

\begin{figure}[htbp]
  \centering
  \includegraphics[width=\columnwidth]{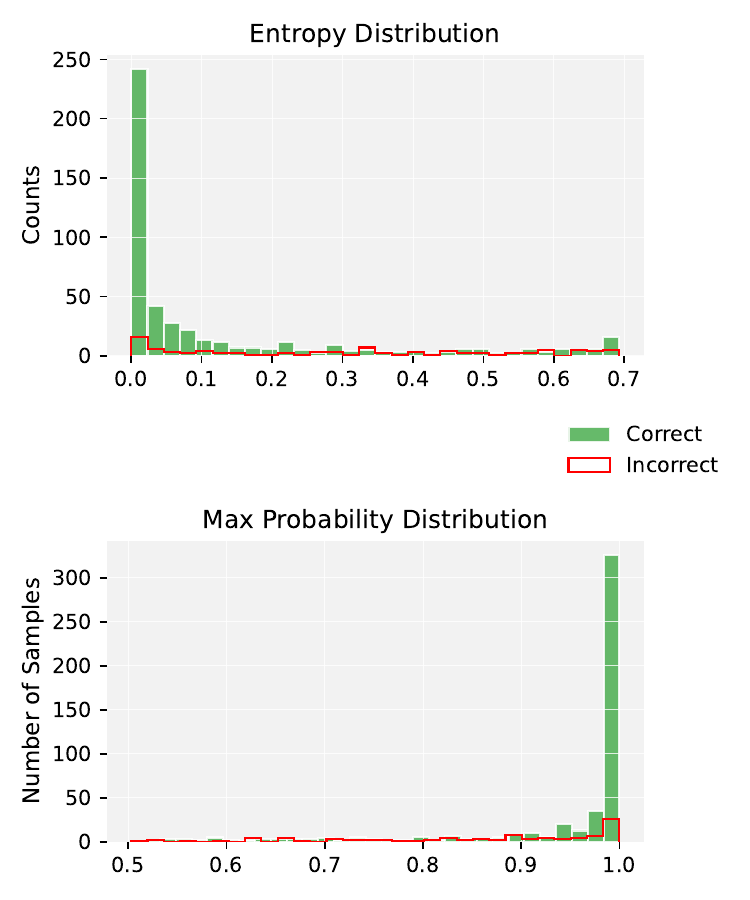}
  \caption{Correct predictions (green) tend to have high confidence and low uncertainty, while incorrect predictions (red) show the opposite pattern. This validates the use of these metrics for high-confidence selection.}
  \label{fig:entropy}
\end{figure}

To ensure a reliable final sample, we selected galaxies with a maximum predicted class probability $p\geq0.99$ and entropy $H\leq0.1$. We chose these thresholds because correctly classified galaxies in the test set usually met these conditions(see Figure~\ref{fig:entropy}). This helps keep only those galaxies the model is confident about, removing uncertain cases. After applying this selection, the precision of class 0 and class 1 in the test dataset increased to 0.95 and 0.89, respectively.

\begin{figure*}[htbp]
  \centering
  \includegraphics[width=\textwidth,height=0.95\textheight,keepaspectratio]{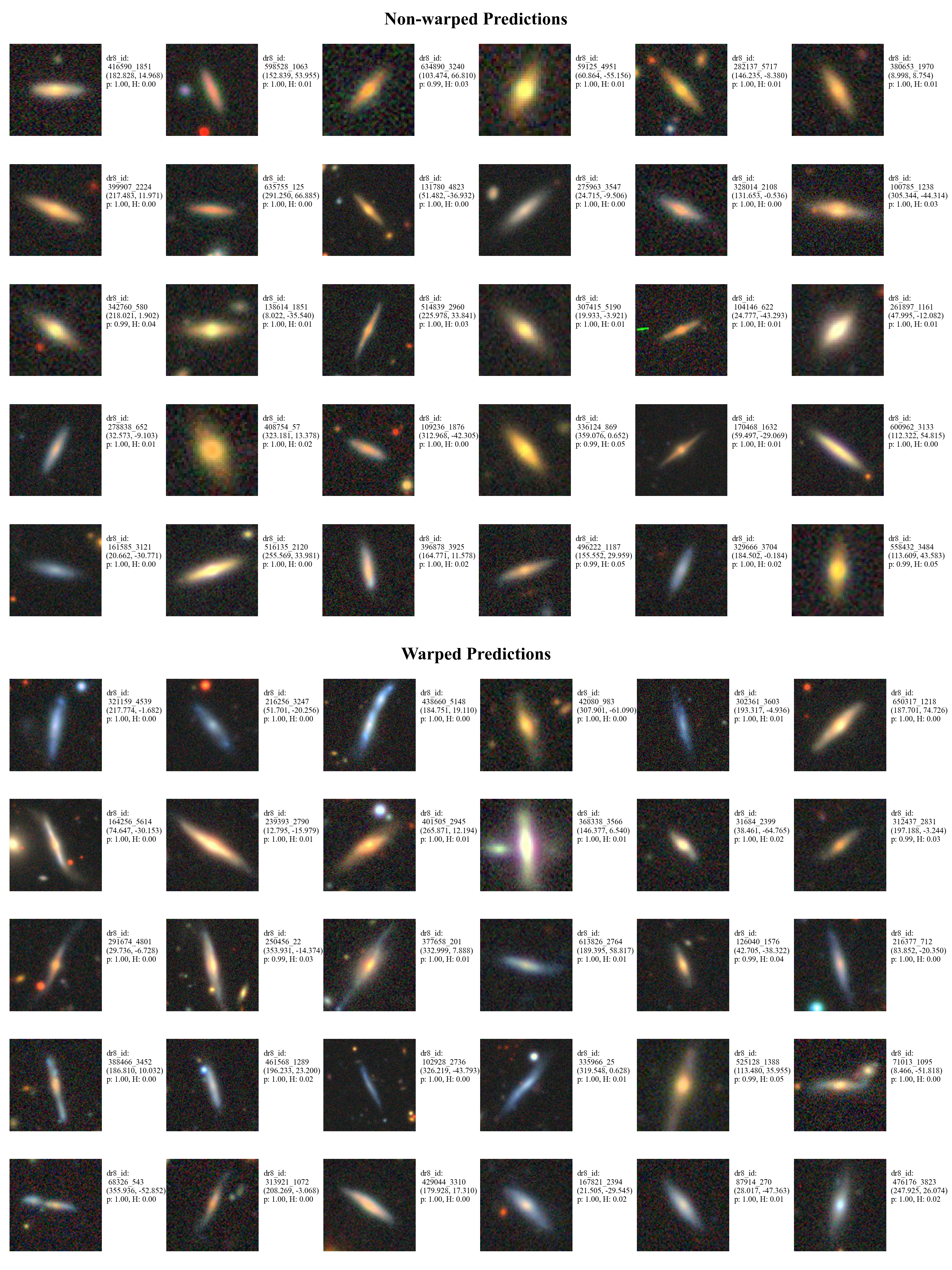}
  \caption{Images of randomly selected galaxies with high-confidence model predictions: the top panel shows non-warped galaxies, and the bottom panel shows warped galaxies.}
  \label{fig:predict_galaxy_images}
\end{figure*}

Using this best-performing model, we applied predictions to 595,651 galaxies that have \texttt{edge-on\_yes\_fraction} greater than or equal to 0.8 from the Galaxy Zoo DESI catalog. Among them, galaxies satisfying the criteria of maximum predicted probability $p\geq$0.99 and entropy $H\leq$0.1 were selected as high-confidence warped and non-warped galaxies. This selection resulted in a final set of 23996 warped and 288562 non-warped edge-on galaxies. A random subset of these high-confidence predictions is shown in Figure \ref{fig:predict_galaxy_images}.

\begin{table}[htbp]
\centering
\tabletypesize{\footnotesize}
\caption{Number of galaxies with valid measurements for morphological and physical properties. }
\begin{tabular}{p{3cm}>{\centering\arraybackslash}p{2.2cm}>{\centering\arraybackslash}p{2.2cm}}
\hline
\multicolumn{3}{l}{\textbf{Morphological Properties}} \\
\hline
Property & Warped Galaxies & Non-warped Galaxies \\
\hline
Sérsic axis ratio & 3372 & 29258 \\
Physical size & 3366 & 29228 \\
Sérsic index & 3372 & 29258 \\
c &  3372 & 29257 \\
\hline
Bulge and merging relevant fractions & 23996 & 288562 \\
\hline
\multicolumn{3}{l}{\textbf{Physical Properties}} \\
\hline
Property & Warped Galaxies & Non-warped Galaxies \\
\hline
$u-g$ & 3372 & 29254 \\
$g-r$ & 3372 & 29256 \\
$r-i$ & 3372 & 29255 \\
$i-z$ & 3371 & 29252 \\
\hline
HI mass & 579 & 727 \\
Stellar Mass & 3372 & 29258 \\
Gas Fraction & 458 & 595 \\
Rotation Speed & 457 & 595 \\
\hline
SFR median, SFR entropy, sSFR median, sSFR entropy & 3111 & 28143 \\
\hline
\end{tabular}
\label{tab:galaxy_counts_combined}
\end{table}

In addition to the main morphology catalog, \citet{walmsley_2023} also released an external catalog, which includes cross-matched properties from other surveys\citep[e.g.,][]{Aguado_2019, haynes_2018, Abazajian_2009, zhou_2021}. Using the \verb|dr8_id|, we matched our classified galaxies to this external catalog and obtained a variety of galaxy properties, including photometry in the $g$, $r$, and $z$ bands, stellar mass, size, axial ratio, Sérsic index, HI mass, star formation rate and other relevant properties. To investigate the environmental context of our galaxies, we performed a cross-match with the SDSS DR7 group catalog \citep{yang_2007} using a 2 arcsecond radius. All galaxies were uniquely matched to a single group, providing a clean sample for subsequent analysis of both intrinsic and group-related properties; approximately 79\% of the galaxies are central members.

Table~\ref{tab:galaxy_counts_combined} summarizes the number of galaxies with valid measurements for each morphological, physical, and star formation property. Based on these measurements, we can now compare the structural properties of warped and non-warped galaxies.

\section{Results}\label{sec:result}

\subsection{Morphological properties}

\begin{figure*}
  \centering
  \includegraphics[width=\textwidth,height=0.95\textheight,keepaspectratio]{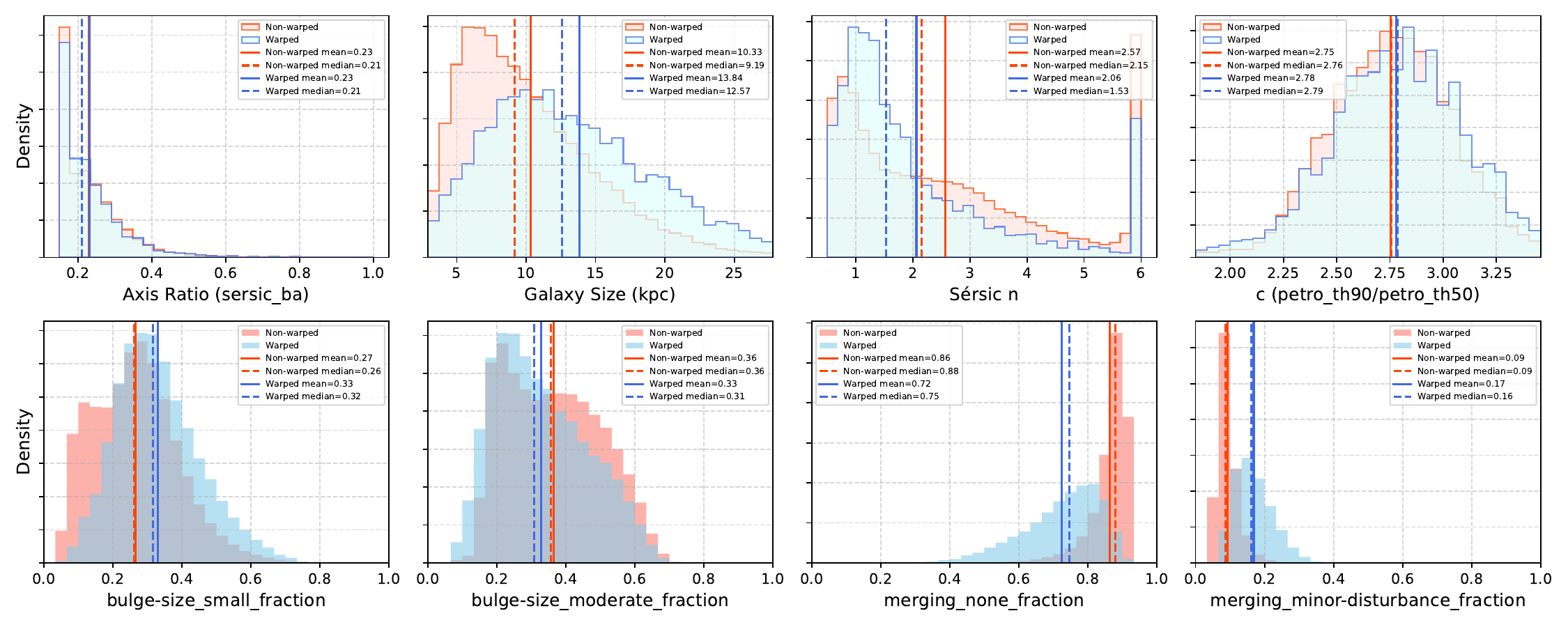}
  \caption{Comparison of morphological properties between warped and non-warped edge-on galaxies. \textbf{Top row:} distributions of Sérsic axis ratio ($b/a$), galaxy size (computed from \texttt{elpetro\_theta\_r}) in physical units (kpc), Sérsic index ($n$), and concentration ($c$, \texttt{petro\_th90}/\texttt{petro\_th50}). Redshifts used for physical size calculation include both spectroscopic and photometric measurements. \textbf{Bottom row:} distributions of bulge size fractions (small and moderate) and merging fractions (none and minor-disturbance). Warped galaxies are shown in blue, non-warped galaxies in red. Vertical solid and dashed lines indicate the mean and median of each sample, respectively. All histograms are normalized to show probability density.}
  \label{fig:morph}
\end{figure*}

We find that the axis ratios (b/a) of warped edge-on disk galaxies are largely comparable to those of non-warped ones (Figure~\ref{fig:morph}, top panel), suggesting that the occurrence of warps is not strongly related to the axis ratio itself. In contrast, warped galaxies exhibit significantly larger effective radii (Figure~\ref{fig:morph}, top panel), which indicates that galaxy size might play a significant role in the occurrence of warps. One possible explanation is that larger galaxies are either intrinsically more susceptible to warp formation due to weaker self-gravity at their outskirts, or that warps in larger systems are easier to detect observationally. Further investigation considering stellar mass, environment, and disk structure is required to disentangle intrinsic physical factors from observational biases. No significant difference between warped and non-warped galaxies is seen in the concentration parameter calculated as \texttt{petro\_th90/petro\_th50} (Figure~\ref{fig:morph}, top panel). This lack of contrast is likely because the warped regions are located in the faint outer parts of the disk, to which \texttt{petro\_th90} is relatively insensitive. 

We also find that warped disk galaxies exhibit systematically lower Sérsic indices compared to non-warped ones (Figure~\ref{fig:morph}, top panel). Although some warped galaxies also show prominent bulges, the overall trend indicates that warps are more common in systems with lower Sérsic indices. The Sérsic index, a measure of the concentration of a galaxy’s light profile, is relatively unaffected by viewing angle in large samples, and therefore serves as a robust proxy for intrinsic morphology.  Lower Sérsic indices are generally associated with disk-dominated galaxies, which are more responsive to gravitational perturbations and thus more likely to exhibit warps. 

Consistent with this view, we examined several morphological predictions from the Galaxy Zoo DESI project, focusing on bulge and merger-related fractions. We find that warped edge-on disk galaxies generally exhibit smaller bulges than their non-warped counterparts, as indicated by systematically higher values in the predicted \texttt{bulge-size\_small\_fraction} and lower values in \texttt{bulge-size\_moderate\_fraction}. These trends suggest that warps predominantly occur in disk-dominated galaxies with relatively small. Warped galaxies also showing lower \texttt{merging-none\_fraction} and higher \texttt{merging\_minor-disturbance\_fraction} (Figure~\ref{fig:morph}, bottom panel). However, caution is required when interpreting the link between warps and minor mergers, since the \texttt{merging\_minor-disturbance\_fraction} is based on visual features. The elevated \texttt{merging\_minor-disturbance\_fraction} among warped galaxies may primarily reflect their morphological irregularities rather than direct evidence of ongoing mergers.

\subsection{physical properties}

\begin{figure*}
  \centering
  \includegraphics[width=\textwidth,height=0.95\textheight,keepaspectratio]{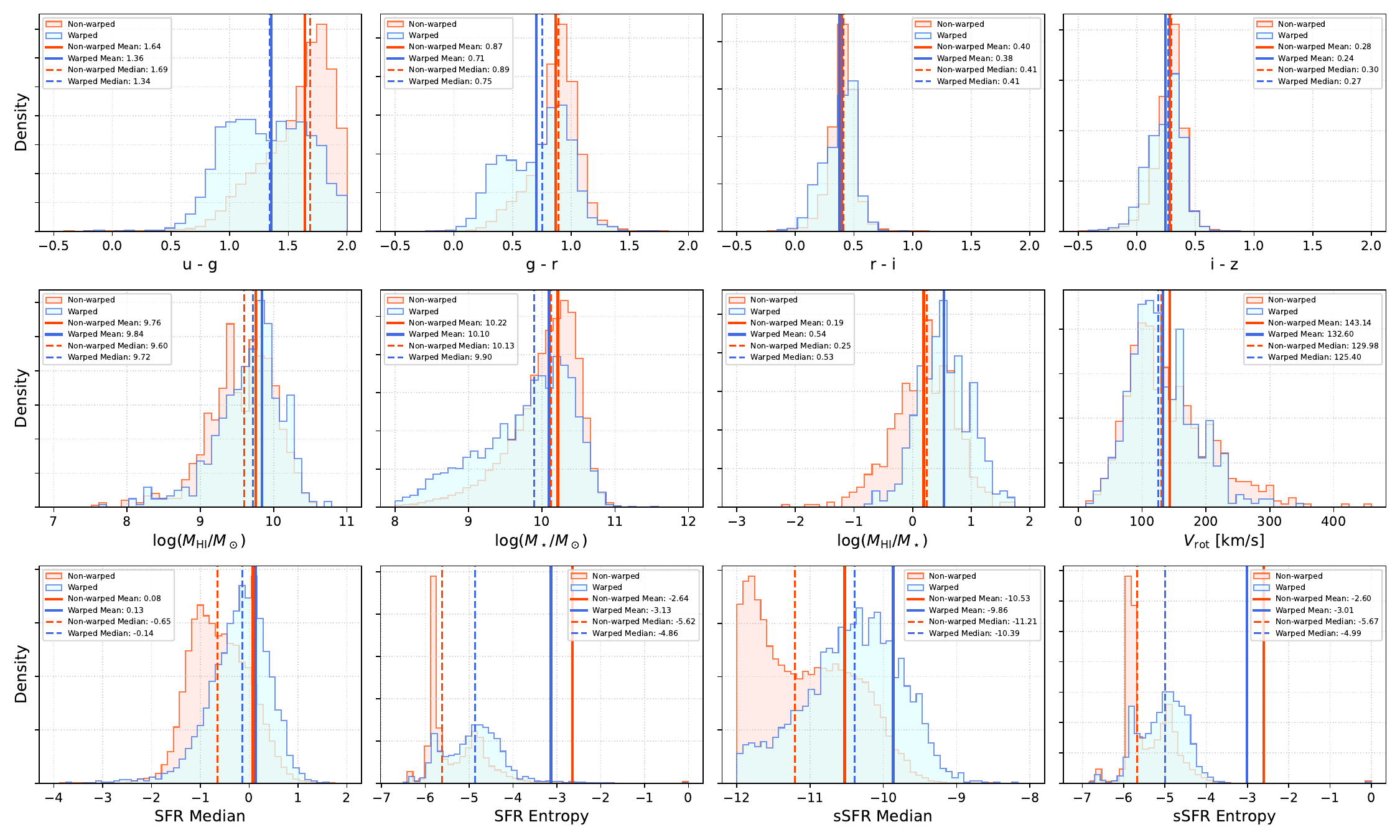}
  \caption{This figure illustrates the differences in physical properties between galaxies. The first row shows the color indices ($u-g$, $g-r$, $r-i$, and $i-z$). The second row presents gas mass, stellar mass, gas fraction, and rotational velocity. The third row displays the median and entropy of star formation rate (SFR) and specific star formation rate (sSFR). Vertical solid and dashed lines indicate the mean and median of each sample, respectively. All histograms are normalized to show probability density.}
  \label{fig:physical}
\end{figure*}

Our result reveals clear physical differences between warped and unwarped galaxies. Overall, warped edge-on disk galaxies tend to be bluer in color, less massive, and more gas-rich(see Figure~\ref{fig:physical}). These characteristics are typically associated with younger, star-forming systems. In contrast, edge-on disk galaxies without warps are generally redder, more massive, and gas-poorer, indicative of more evolved systems with lower star formation activity.

On average, non-warped galaxies are redder than warped galaxies, with higher $u - g$ and $g - r$ values. The differences in $r - i$ and in $i - z$ are almost negligible (Figure~\ref{fig:physical}, top panel). In general, the color contrast is more pronounced at shorter wavelengths, as seen in the larger difference in $u - g$ and $g - r$ compared to $r - i$ and $i - z$. This indicates that the observed differences primarily reflect variations in the light from relatively younger stellar populations, while the contributions from older stars traced by longer-wavelength colors are similar. While these color differences alone cannot unambiguously indicate the current star formation activity, warped galaxies do exhibit higher star formation rates, as illustrated in the last panel of Figure~\ref{fig:physical}.

Among disk galaxies successfully cross-matched with ALFALFA observations, we compared the HI gas content of warped and non-warped systems. We find that warped galaxies exhibit slightly higher HI masses compared to their non-warped counterparts, while their stellar masses are significantly lower. As a result, warped galaxies have systematically higher gas fractions. This trend suggests that warped galaxies are generally more gas-rich and less evolved on average in terms of stellar mass assembly (Figure~\ref{fig:physical}, middle panel).
It is worth noting, however, that the ALFALFA survey is subject to strong selection effects: only galaxies with sufficiently high HI content can be detected. Therefore, the non-warped galaxies included in this comparison represent the gas-rich, star-forming population, rather than the entire non-warped sample. Consequently, the difference in gas fraction between warped and non-warped galaxies may be even more pronounced when considering the full galaxy population. We also compared the overall probability distribution functions of rotational velocities for warped and non-warped galaxies, and found the distributions to be comparable. The rotational velocities were estimated from HI line widths corrected for inclination using Sersic axis ratios, so the accuracy for individual galaxies is limited.

We further investigated the star formation properties of warped versus non-warped galaxies and found that warped galaxies exhibit systematically higher star formation rates (SFR), specific star formation rates (sSFR), and higher SFR entropy. The PDF of log(sSFR) for edge-on galaxies in the non-warped sample exhibits a clear bimodal distribution, with one peak around typical star-forming values and another significant peak at log(sSFR)$<$ $-11$ (Figure~\ref{fig:physical}, bottom panel). This low-sSFR peak indicates that a fraction of these galaxies have already quenched, likely transitioning into a quiescent or passive state, whereas the majority of warped edge-on galaxies remain actively star-forming. Also, the SFR/sSFR entropy values of warped galaxies are generally higher, indicating a broader distribution of star formation activity, whereas non-warped galaxies show a pronounced peak at relatively low entropy values.

\subsection{Dependence on Galaxy Masses}

\begin{figure*}
  \centering
  \includegraphics[width=\textwidth,height=0.95\textheight,keepaspectratio]{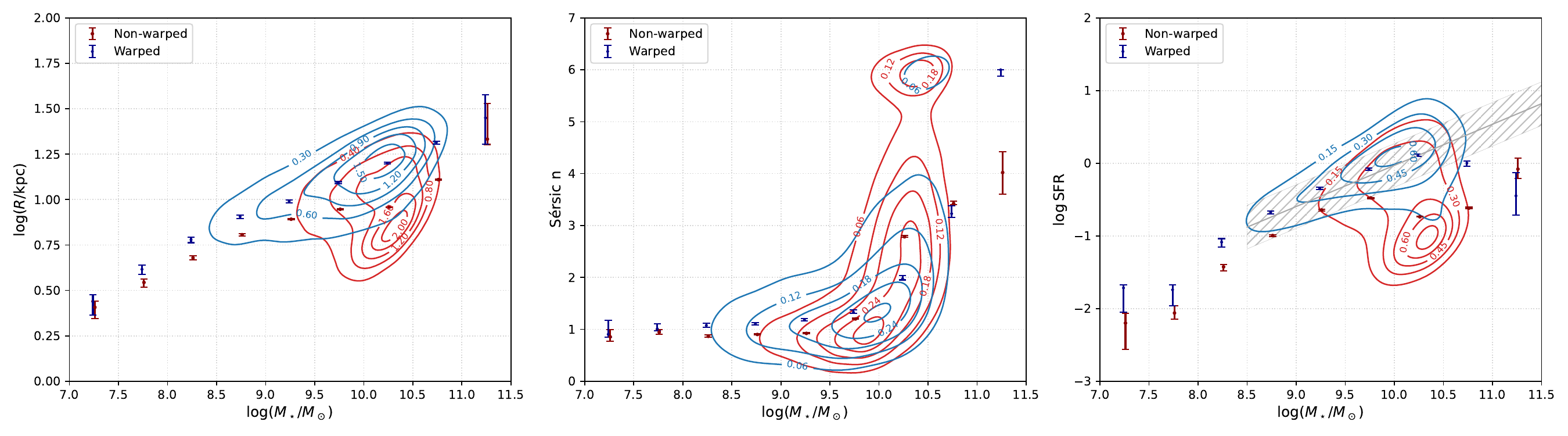}
  \caption{Density distributions of key physical and structural properties as a function of stellar mass for warped (blue) and non-warped (red) disk galaxies. Each panel shows kernel density contours of the two populations, with binned medians and corresponding $1\sigma$ uncertainties overplotted as symbols. From left to right and top to bottom, the panels show: (a) Disk Size $R$ (kpc), (b) Sérsic index $n$, (c) star-formation rate (SFR). The gray line in panel (c) denotes the star-forming main sequence from \citet{speagle_2014}, with the shaded region indicating its $\pm0.3$~dex intrinsic scatter.}
  \label{fig:dependence_on_mass}
\end{figure*}

Many galaxy properties are strongly correlated with stellar mass. To ensure that the differences we observe between warped and non-warped galaxies reflect the impact of warps rather than underlying mass trends, we control for stellar mass in our analysis, by comparing galaxies within similar stellar mass bins.

As have been shown earlier, warped galaxies have, on average, larger physical sizes (Figure~\ref{fig:morph}). However, their stellar masses are generally lower(Figure~\ref{fig:physical}). The 2D mass–size distribution reveals that warped galaxies are more widely dispersed across the plane than non-warped ones; within the same mass range, they tend to have larger sizes(Figure~\ref{fig:dependence_on_mass}). Dividing the sample into stellar mass bins, we further found that warped galaxies exhibit larger median sizes than non-warped galaxies across all mass bins, and the size difference between warped and non-warped galaxies becomes more pronounced at higher stellar masses. This trend supports that the occurrence of warps is likely related to the structure of the outer stellar disk, where more extended or less tightly bound outer regions could be more susceptible to warping. In contrast, many high-mass non-warped galaxies may be bulge-dominated, which contributes to the observed size difference.

Figure~\ref{fig:dependence_on_mass} shows that warped galaxies tend to exhibit lower Sérsic indices at fixed stellar mass compared to non-warped counterparts at intermediate masses (\(\log(M_\star/M_\odot) \sim 10.0 - 11.0\)), suggesting that warped systems are generally more disk-dominated. At the low- and high-mass ends, the differences between the two populations are less pronounced, and in some cases the trend even appears to reverse. Overall, this trend is consistent with warped galaxies preferentially residing in late-type, disk-dominated systems.

The 2D distribution on the stellar mass–SFR plane (Figure~\ref{fig:dependence_on_mass}) shows that at all stellar mass ranges, the median SFR of warped galaxies remains higher than that of non-warped galaxies. The result further supports the trend in sSFR distribution(Figure~\ref{fig:physical}) : warped edge-on galaxies rarely appear in the quenched region and are predominantly located along, or slightly above, the star-forming main sequence, exhibiting systematically higher SFRs than both the main sequence average and their non-warped counterparts. Notably, quenched galaxies are almost entirely absent from the warped population. In contrast, non-warped galaxies show a broader spread, with a significant fraction lying below the main sequence—indicative of quenching activity.

\subsection{Environmental Difference}

\begin{figure}
  \centering
  \includegraphics[width=\columnwidth]{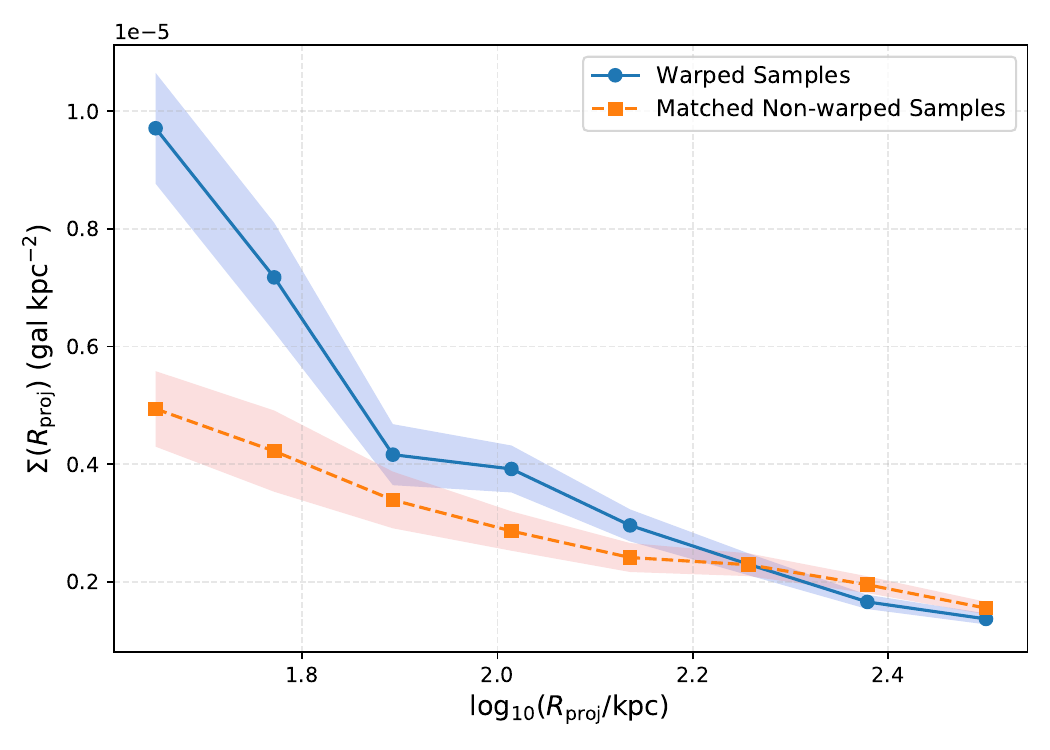}
  \caption{Projected neighbor surface density profiles, $\Sigma(R_{\mathrm{proj}})$, for warped (blue) and matched non-warped (red) galaxies as a function of projected separation $R_{\mathrm{proj}}$ (in kpc). Solid circles and blue shading denote the mean and standard error of the mean (SEM) for warped galaxies, while dashed squares and red shading show the corresponding values for matched non-warped control samples. Warped galaxies exhibit systematically higher local projected densities at small separations.}
  \label{fig:proj_density}
\end{figure}

To minimize the influence of other confounding factors, we applied a matched-control strategy to the warped galaxies by searching for their nearest non-warped counterparts in the parameter space of stellar mass, color, and redshift.
Given that the ratio between high-confidence warped and non-warped galaxies is approximately 1:10, we selected seven non-warped counterparts for each warped galaxy that closely match in stellar mass, color, and redshift.
Since only galaxies within the SDSS survey volume have reliable stellar mass measurements, we further cross-matched our sample with the SDSS DR7 group catalog from \citet{yang_2007} to investigate their environmental properties.
For each galaxy, a counterpart was identified within a two arcsecond matching radius.

We quantify the local environmental densities around warped galaxies and their non-warped matched counterparts by computing the \textit{projected surface number density} of neighboring galaxies within concentric annuli, defined as
\[
\Sigma(R_{\mathrm{proj}}) =
\frac{N(R_i < R_{\mathrm{proj}} < R_{i+1})}
{\pi \left(R_{i+1}^2 - R_i^2\right)},
\]
where \( R_{\mathrm{proj}} \) denotes the projected separation from the target galaxy on the sky. 
This annular measurement traces the spatial distribution of neighboring galaxies as a function of scale. To select neighboring galaxies within each annulus, we apply a line-of-sight constraint based on redshift. The maximum allowed redshift difference is  
\[
\Delta z_\mathrm{max} = \frac{\Delta v_\mathrm{max}}{c} \,(1 + z_\mathrm{target}),
\]  
where \(\Delta v_\mathrm{max} = 500~\mathrm{km\,s^{-1}}\) and \(c\) is the speed of light. Only galaxies with 
\(|z_\mathrm{neighbor} - z_\mathrm{target}| \le \Delta z_\mathrm{max}\) are counted as neighbors, minimizing contamination from foreground or background objects along the line of sight.

As shown in Figure~\ref{fig:proj_density}, warped galaxies (blue) tend to have higher annular surface densities than their matched non-warped counterparts (red) at small to intermediate projected radii. The difference is especially pronounced within 
\(\log_{10}(R_{\mathrm{proj}}/\mathrm{kpc}) \lesssim 1.7\)
(\(R_{\mathrm{proj}} \lesssim 50~\mathrm{kpc}\)), 
where the local number density around warped galaxies is roughly twice as high as that around non-warped galaxies. This trend indicates an association between disk warping and denser immediate environments, although it is unclear whether the environment directly induces warps or merely traces other galaxy properties linked to warping. At larger projected separations 
(\(R_{\mathrm{proj}} \gtrsim 200~\mathrm{kpc}\)), 
the two profiles converge, implying that any connection between disk warping and environment is more likely limited to the local, small-scale surroundings. We note, however, that these measurements only consider the projected surface density of neighboring galaxies; more detailed studies, including kinematic information and three-dimensional environmental reconstruction, will be required to establish whether local interactions play a causal role in inducing disk warps.

\section{Summary}\label{sec:diss}
By applying a convolutional neural network classifier to edge-on disk galaxy images from the DESI DR8 Legacy Imaging Surveys, we constructed the largest catalog of warped galaxies to date, enabling statistical comparisons between warped and non-warped systems for the first time. This work is based solely on optical imaging data, with the sample restricted to edge-on disk galaxies to ensure reliable identification of warps. The result reveals that these warped edge-on galaxies exhibit notable morphological, physical, and environmental differences compared to their non-warped counterparts. 

Morphologically, warped galaxies tend to have larger disks, lower bulge-relevant fractions, and higher merging-relevant fractions, consistent with late-type, disk-dominated structures that may be more susceptible to warping. Their axis ratios are comparable to those of non-warped galaxies, indicating similar overall flattening. When dividing galaxies into stellar mass bins, warped galaxies exhibit larger median sizes than non-warped galaxies across all bins, with the difference becoming more pronounced at higher stellar masses. The trend indicates that warps may be associated with the structure of the outer stellar disk, where extended or loosely bound outer regions could be more prone to warping, whereas high-mass non-warped galaxies tend to be bulge-dominated, contributing to the size disparity. Overall, non-warped galaxies exhibit higher Sérsic indices than warped ones; this trend is primarily driven by the intermediate-mass range, where a pronounced peak around $n\sim6$ reflects a population of bulge-dominated systems. At lower and higher masses, by contrast, warped galaxies tend to show comparable or even slightly higher Sérsic indices.

Physically, warped galaxies tend to be less massive, bluer, and more gas-rich than their non-warped counterparts. The color difference between warped and non-warped galaxies is particularly pronounced at shorter wavelengths, consistent with their higher SFRs, reflecting more active star formation in warped galaxies, while at the longest wavelengths the two populations show little color distinction. On the stellar mass–SFR plane, warped galaxies largely follow the star-forming main sequence, often lying slightly above it, indicative of active star formation and mildly elevated SFRs compared to the average star-forming population. In contrast, many non-warped galaxies fall below this sequence, suggesting they are quenched or transitioning into quiescence, a distinction reinforced by the sSFR distribution, where non-warped galaxies exhibit a bimodal pattern with a prominent low-sSFR peak below log(sSFR) $\approx -11$, characteristic of quenched systems. It should be noted that our sample selection is based solely on the edge-on criterion \texttt{edge-on\_yes\_fraction}, and due to projection effects it is difficult to identify the presence of spiral arms in these systems; thus, the potential contribution of S0 galaxies with lower star formation cannot be fully excluded. Overall, warped galaxies generally appear to represent an earlier evolutionary stage compared to non-warped galaxies.

In terms of environmental properties, we find that warped galaxies tend to have a higher number of nearby companions than their non-warped counterparts, particularly within $R_{\mathrm{proj}} \lesssim 50~\mathrm{kpc}$, where the local projected number density is roughly twice as high. At larger separations, the environmental difference disappears, indicating that warps preferentially reside in locally denser regions, although it remains unclear whether this reflects a causal effect or merely traces other environmental properties.

Together, these results suggest that disk warping is closely linked to the structural properties, star formation activity, and small-scale environments of galaxies, highlighting the interplay between internal dynamics and local interactions. Warps preferentially occur in late-type disk galaxies with larger galaxy sizes and less massive bulges, which are also more gas-rich and actively star-forming, and may be more susceptible to external torques or internal dynamical instabilities. They are more often found in denser local environments at small separations, suggesting a possible link between warps and interactions. However, establishing a direct causal connection requires more detailed observations, including stellar and gas kinematics, three-dimensional reconstructions of local environments, and deep imaging of faint outer disks. High-resolution simulations of galaxy interactions and gas accretion will also be essential to understand the physical mechanisms driving warp formation and their evolution over cosmic time.


\begin{acknowledgments}
This work made use of the publicly available data from the Galaxy Zoo DESI project, the DESI Legacy Imaging Surveys DR8, and the SDSS DR7 group catalog constructed and released by Xiaohu Yang and collaborators. We would like to thank the Galaxy Zoo DESI team, the DESI Legacy Imaging Surveys team, and the Shanghai Jiao Tong University group catalog team for making their data products publicly available.

YHW thanks Houzun Chen and Ziyong Wu for helpful discussions and suggestions.

This work is supported by the National Key Research and Development Program (No. 2022YFA1602903, No. 2023YFB3002502), NSFC(No.12533007), the science research grants from the China Manned Space project with NO.CMS-CSST-2025-A10. JFL thanks research grants from the National Natural Science Foundation of China (No. 12403016) and the Postdoctoral Fellowship Program of CPSF (No. GZC20241514). HYW is supported by the National Natural Science Foundation of China (NSFC, Nos. 12192224). EW thanks support of the National Science Foundation of China (Nos. 12473008). 
\end{acknowledgments}

\appendix
\twocolumngrid
\section{View Selection Criteria}\label{app:view}

\begin{figure}[htbp]
  \centering
  \includegraphics[width=0.5\textwidth]{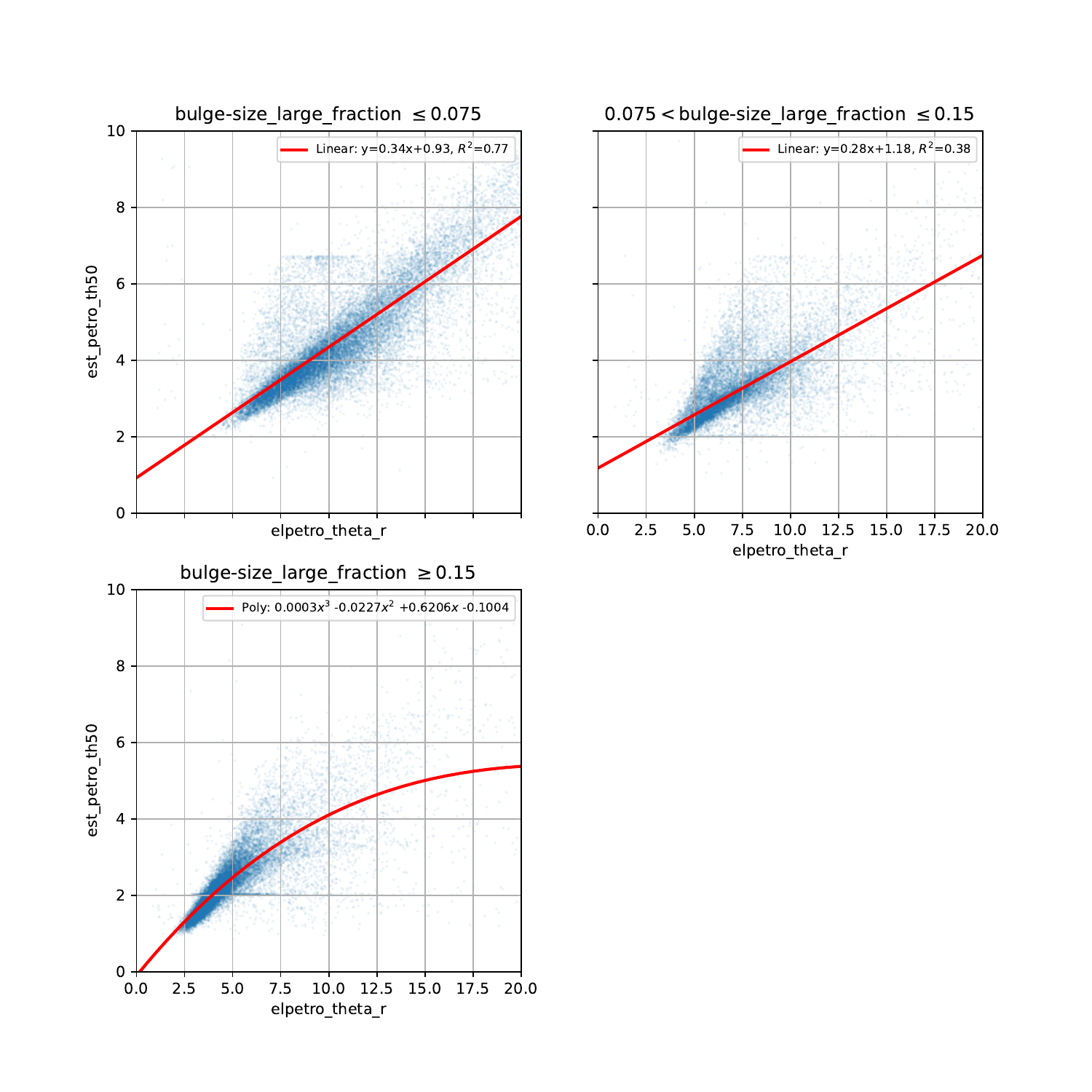}
  \caption{Correlation between estimated Petrosian half-light radius (\texttt{est\_petro\_th50}) and Petrosian angular size (\texttt{elpetro\_theta\_r}) for galaxies in different bulge-dominated regimes. The sample is divided into three bins of \texttt{bulge-size\_large\_fraction}: $\leq 0.075$ (top left), $0.075 < $fraction$\leq 0.15$ (top right), and $\geq 0.15$ (bottom). Scatter points represent individual galaxies, while red curves show the best-fit relations. The correlation is tightest for the lowest bulge fractions, weakens for intermediate fractions, and becomes nonlinear for the most bulge-dominated systems.}
  \label{fig:viewselect}
\end{figure}

To ensure that each cutout image captures the full structure of a galaxy while remaining focused on its morphology, an appropriate cutout size must be selected for each galaxy. We found that relying solely on the estimated Petrosian half-light radius (\texttt{est\_petro\_th50}) does not provide suitable sizes for all galaxies, particularly for edge-on systems with prominent bulges, where the galaxies' angular sizes tend to be underestimated. While \texttt{elpetro\_theta\_r} offers a more reliable measure for setting the image scale, some galaxies in the Galaxy Zoo DESI catalog lack valid \texttt{elpetro\_theta\_r} values. For these cases, we estimate \texttt{elpetro\_theta\_r} from their available \texttt{est\_petro\_th50} values, using an empirical relation derived from galaxies with both parameters measured.

We first investigated the relationship between the estimated Petrosian half-light radius (\texttt{est\_petro\_th50}) and the elliptical Petrosian radius (\texttt{elpetro\_theta\_r}). Galaxies were initially divided into three morphological subgroups according to their \texttt{bulge-size\_large\_fraction}, corresponding to disk-dominated, intermediate, and bulge-dominated systems. As illustrated in Fig.~\ref{fig:viewselect}, the correlations between \texttt{est\_petro\_th50} and \texttt{elpetro\_theta\_r} differ across these subgroups: a robust linear relation is found for disk-dominated galaxies, a weaker and more scattered relation for intermediate systems, and a nonlinear trend better captured by a third-order polynomial for bulge-dominated galaxies. These empirical fits provide a basis for estimating \texttt{elpetro\_theta\_r} in cases where it is missing in the catalog.

For galaxies with \texttt{bulge-size\_large\_fraction} $\leq 0.15$, we set

\[
\theta_r = 2 \times \texttt{est\_petro\_th50}
\]

while for galaxies with larger bulges (\texttt{bulge-size\_large\_fraction} $> 0.15$), we estimate $\theta_r$ by inverting the polynomial relation calibrated in Fig.~\ref{fig:viewselect} when $\texttt{est\_petro\_th50} \lesssim 5.6$, and apply fixed upper bounds of $\theta_r = 25$ or $\theta_r = 30$ for more extended cases. The final cutout size is then defined as
\[
\text{cutout\_size} = \text{round}\left(\frac{4 \times \theta_r}{\text{PIX\_SCALE}}\right)
\]

Nevertheless, we found that in the intermediate bulge regime, a subset of galaxies exhibits significantly overestimated values of \texttt{elpetro\_theta\_r}, which would result in excessively large image cutouts and hinder the subsequent machine-learning classification. To mitigate this, we adopted an alternative parameterization based on \texttt{bulge-size\_small\_fraction}, which provides a more practical morphological separation for our purpose. 

\section{Training vs. Predicted Sample Properties}\label{app:train_predict}
\begin{figure*}
  \centering
  \includegraphics[width=0.9\textwidth,height=0.9\textheight,keepaspectratio]{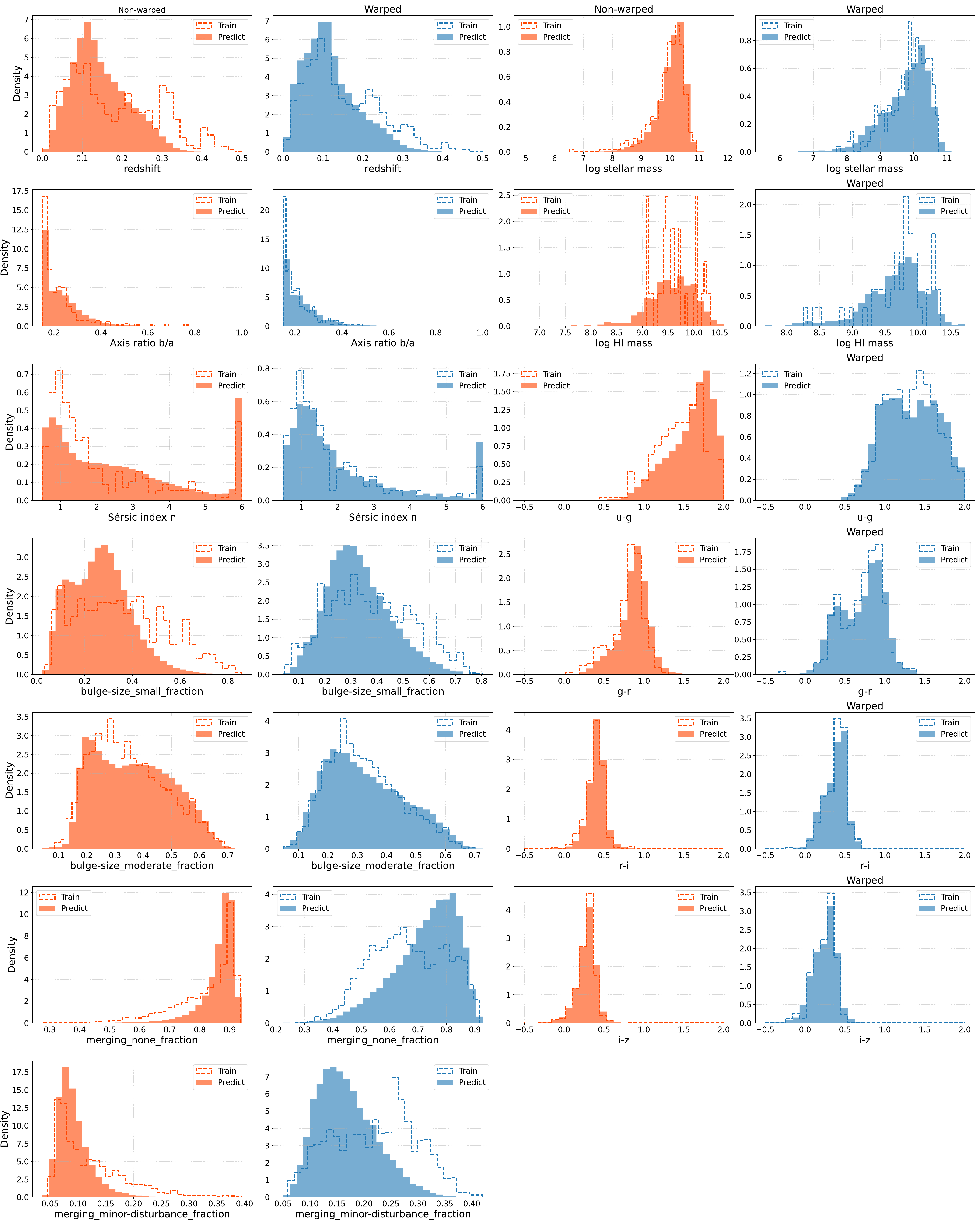}
  \caption{Comparison of distributions for various physical properties between the training and predicted samples of warped and non-warped galaxies. The left half (first two columns) shows seven physical quantities: redshift, axis ratio b/a, Sérsic index n, bulge fraction, bulge moderate fraction, merging none fraction, and merging minor disturbance fraction. The right half (last two columns) shows the remaining six quantities: log stellar mass, log HI mass, and four colors ($u-g, g-r, r-i, i-z$). Each row corresponds to a specific quantity, with the first column for non-warped galaxies and the second column for warped galaxies. Dashed lines represent the training sample, while filled histograms represent the predicted sample.}
  \label{fig:predict_vs_train}
\end{figure*}

To verify the representativeness and reliability of our trained classifier, we compare the distributions of various physical properties between the original training samples and the predicted samples for both warped and non-warped galaxies (See in Figure\ref{fig:predict_vs_train}). This comparison allows us to check whether the predicted samples preserve the overall characteristics of the training data, ensuring that the classifier does not introduce significant biases in terms of redshift, morphology, stellar mass, gas content, or colors. By examining these distributions, we can confirm that the predicted samples remain consistent with the physical trends observed in the training set, providing confidence that the classifier produces physically meaningful results across the full galaxy sample.

\bibliography{reference}{}

@article{walmsley_2021,
	title = {Galaxy {Zoo} {DECaLS}: {Detailed} visual morphology measurements from volunteers and deep learning for 314 000 galaxies},
	volume = {509},
	copyright = {https://creativecommons.org/licenses/by/4.0/},
	issn = {0035-8711, 1365-2966},
	shorttitle = {Galaxy {Zoo} {DECaLS}},
	url = {https://academic.oup.com/mnras/article/509/3/3966/6378289},
	doi = {10.1093/mnras/stab2093},
	language = {en},
	number = {3},
	urldate = {2025-05-15},
	journal = {Monthly Notices of the Royal Astronomical Society},
	author = {Walmsley, Mike and Lintott, Chris and Géron, Tobias and Kruk, Sandor and Krawczyk, Coleman and Willett, Kyle W and Bamford, Steven and Kelvin, Lee S and Fortson, Lucy and Gal, Yarin and Keel, William and Masters, Karen L and Mehta, Vihang and Simmons, Brooke D and Smethurst, Rebecca and Smith, Lewis and Baeten, Elisabeth M and Macmillan, Christine},
	month = dec,
	year = {2021},
	pages = {3966--3988},
}

@article{walmsley_2023,
	title = {Galaxy {Zoo} {DESI}: {Detailed} morphology measurements for 8.{7M} galaxies in the {DESI} {Legacy} {Imaging} {Surveys}},
	volume = {526},
	copyright = {https://creativecommons.org/licenses/by/4.0/},
	issn = {0035-8711, 1365-2966},
	shorttitle = {Galaxy {Zoo} {DESI}},
	url = {https://academic.oup.com/mnras/article/526/3/4768/7283169},
	doi = {10.1093/mnras/stad2919},
	language = {en},
	number = {3},
	urldate = {2025-03-03},
	journal = {Monthly Notices of the Royal Astronomical Society},
	author = {Walmsley, Mike and Géron, Tobias and Kruk, Sandor and Scaife, Anna M M and Lintott, Chris and Masters, Karen L and Dawson, James M and Dickinson, Hugh and Fortson, Lucy and Garland, Izzy L and Mantha, Kameswara and O’Ryan, David and Popp, Jürgen and Simmons, Brooke and Baeten, Elisabeth M and Macmillan, Christine},
	month = oct,
	year = {2023},
	pages = {4768--4786},
}

@misc{Reshetnikov_2025,
	title = {Galactic warps: from cosmic noon to the current epoch},
	shorttitle = {Galactic warps},
	url = {http://arxiv.org/abs/2504.12403},
	doi = {10.48550/arXiv.2504.12403},
	language = {en},
	urldate = {2025-05-07},
	publisher = {arXiv},
	author = {Reshetnikov, Vladimir P. and Chugunov, Ilia V. and Marchuk, Alexander A. and Mosenkov, Aleksandr V. and Kozlov, Matvey D. and Savchenko, Sergey S. and Makarov, Dmitry I. and Antipova, Aleksandra V. and Sypkova, Anastasia M.},
	month = apr,
	year = {2025},
	note = {arXiv:2504.12403 [astro-ph]},
}

@article{dey_overview_2019,
	title = {Overview of the {DESI} {Legacy} {Imaging} {Surveys}},
	volume = {157},
	issn = {0004-6256, 1538-3881},
	url = {https://iopscience.iop.org/article/10.3847/1538-3881/ab089d},
	doi = {10.3847/1538-3881/ab089d},
	language = {en},
	number = {5},
	urldate = {2025-05-18},
	journal = {The Astronomical Journal},
	author = {Dey, Arjun and Schlegel, David J. and Lang, Dustin and Blum, Robert and Burleigh, Kaylan and Fan, Xiaohui and Findlay, Joseph R. and Finkbeiner, Doug and Herrera, David and Juneau, Stéphanie and Landriau, Martin and Levi, Michael and McGreer, Ian and Meisner, Aaron and Myers, Adam D. and Moustakas, John and Nugent, Peter and Patej, Anna and Schlafly, Edward F. and Walker, Alistair R. and Valdes, Francisco and Weaver, Benjamin A. and Yèche, Christophe and Zou, Hu and Zhou, Xu and Abareshi, Behzad and Abbott, T. M. C. and Abolfathi, Bela and Aguilera, C. and Alam, Shadab and Allen, Lori and Alvarez, A. and Annis, James and Ansarinejad, Behzad and Aubert, Marie and Beechert, Jacqueline and Bell, Eric F. and BenZvi, Segev Y. and Beutler, Florian and Bielby, Richard M. and Bolton, Adam S. and Briceño, César and Buckley-Geer, Elizabeth J. and Butler, Karen and Calamida, Annalisa and Carlberg, Raymond G. and Carter, Paul and Casas, Ricard and Castander, Francisco J. and Choi, Yumi and Comparat, Johan and Cukanovaite, Elena and Delubac, Timothée and DeVries, Kaitlin and Dey, Sharmila and Dhungana, Govinda and Dickinson, Mark and Ding, Zhejie and Donaldson, John B. and Duan, Yutong and Duckworth, Christopher J. and Eftekharzadeh, Sarah and Eisenstein, Daniel J. and Etourneau, Thomas and Fagrelius, Parker A. and Farihi, Jay and Fitzpatrick, Mike and Font-Ribera, Andreu and Fulmer, Leah and Gänsicke, Boris T. and Gaztanaga, Enrique and George, Koshy and Gerdes, David W. and A Gontcho, Satya Gontcho and Gorgoni, Claudio and Green, Gregory and Guy, Julien and Harmer, Diane and Hernandez, M. and Honscheid, Klaus and Huang, Lijuan (Wendy) and James, David J. and Jannuzi, Buell T. and Jiang, Linhua and Joyce, Richard and Karcher, Armin and Karkar, Sonia and Kehoe, Robert and Kneib, , Jean-Paul and Kueter-Young, Andrea and Lan, Ting-Wen and Lauer, Tod R. and Guillou, Laurent Le and Van Suu, Auguste Le and Lee, Jae Hyeon and Lesser, Michael and Levasseur, Laurence Perreault and Li, Ting S. and Mann, Justin L. and Marshall, Robert and Martínez-Vázquez, C. E. and Martini, Paul and Du Mas Des Bourboux, Hélion and McManus, Sean and Meier, Tobias Gabriel and Ménard, Brice and Metcalfe, Nigel and Muñoz-Gutiérrez, Andrea and Najita, Joan and Napier, Kevin and Narayan, Gautham and Newman, Jeffrey A. and Nie, Jundan and Nord, Brian and Norman, Dara J. and Olsen, Knut A. G. and Paat, Anthony and Palanque-Delabrouille, Nathalie and Peng, Xiyan and Poppett, Claire L. and Poremba, Megan R. and Prakash, Abhishek and Rabinowitz, David and Raichoor, Anand and Rezaie, Mehdi and Robertson, A. N. and Roe, Natalie A. and Ross, Ashley J. and Ross, Nicholas P. and Rudnick, Gregory and Gaines, Sasha and Saha, Abhijit and Sánchez, F. Javier and Savary, Elodie and Schweiker, Heidi and Scott, Adam and Seo, Hee-Jong and Shan, Huanyuan and Silva, David R. and Slepian, Zachary and Soto, Christian and Sprayberry, David and Staten, Ryan and Stillman, Coley M. and Stupak, Robert J. and Summers, David L. and Tie, Suk Sien and Tirado, H. and Vargas-Magaña, Mariana and Vivas, A. Katherina and Wechsler, Risa H. and Williams, Doug and Yang, Jinyi and Yang, Qian and Yapici, Tolga and Zaritsky, Dennis and Zenteno, A. and Zhang, Kai and Zhang, Tianmeng and Zhou, Rongpu and Zhou, Zhimin},
	month = may,
	year = {2019},
	pages = {168},
}

@article{flaugher_dark_2015,
	title = {{THE} {DARK} {ENERGY} {CAMERA}},
	volume = {150},
	copyright = {http://iopscience.iop.org/info/page/text-and-data-mining},
	issn = {1538-3881},
	url = {https://iopscience.iop.org/article/10.1088/0004-6256/150/5/150},
	doi = {10.1088/0004-6256/150/5/150},
	language = {en},
	number = {5},
	urldate = {2025-05-19},
	journal = {The Astronomical Journal},
	author = {Flaugher, B. and Diehl, H. T. and Honscheid, K. and Abbott, T. M. C. and Alvarez, O. and Angstadt, R. and Annis, J. T. and Antonik, M. and Ballester, O. and Beaufore, L. and Bernstein, G. M. and Bernstein, R. A. and Bigelow, B. and Bonati, M. and Boprie, D. and Brooks, D. and Buckley-Geer, E. J. and Campa, J. and Cardiel-Sas, L. and Castander, F. J. and Castilla, J. and Cease, H. and Cela-Ruiz, J. M. and Chappa, S. and Chi, E. and Cooper, C. and Da Costa, L. N. and Dede, E. and Derylo, G. and DePoy, D. L. and De Vicente, J. and Doel, P. and Drlica-Wagner, A. and Eiting, J. and Elliott, A. E. and Emes, J. and Estrada, J. and Fausti Neto, A. and Finley, D. A. and Flores, R. and Frieman, J. and Gerdes, D. and Gladders, M. D. and Gregory, B. and Gutierrez, G. R. and Hao, J. and Holland, S. E. and Holm, S. and Huffman, D. and Jackson, C. and James, D. J. and Jonas, M. and Karcher, A. and Karliner, I. and Kent, S. and Kessler, R. and Kozlovsky, M. and Kron, R. G. and Kubik, D. and Kuehn, K. and Kuhlmann, S. and Kuk, K. and Lahav, O. and Lathrop, A. and Lee, J. and Levi, M. E. and Lewis, P. and Li, T. S. and Mandrichenko, I. and Marshall, J. L. and Martinez, G. and Merritt, K. W. and Miquel, R. and Muñoz, F. and Neilsen, E. H. and Nichol, R. C. and Nord, B. and Ogando, R. and Olsen, J. and Palaio, N. and Patton, K. and Peoples, J. and Plazas, A. A. and Rauch, J. and Reil, K. and Rheault, J.-P. and Roe, N. A. and Rogers, H. and Roodman, A. and Sanchez, E. and Scarpine, V. and Schindler, R. H. and Schmidt, R. and Schmitt, R. and Schubnell, M. and Schultz, K. and Schurter, P. and Scott, L. and Serrano, S. and Shaw, T. M. and Smith, R. C. and Soares-Santos, M. and Stefanik, A. and Stuermer, W. and Suchyta, E. and Sypniewski, A. and Tarle, G. and Thaler, J. and Tighe, R. and Tran, C. and Tucker, D. and Walker, A. R. and Wang, G. and Watson, M. and Weaverdyck, C. and Wester, W. and Woods, R. and Yanny, B. and {The DES Collaboration}},
	month = oct,
	year = {2015},
	pages = {150},
}

@inproceedings{tan_19a,
  title     = {EfficientNet: Rethinking Model Scaling for Convolutional Neural Networks},
  author    = {Tan, Mingxing and Le, Quoc V.},
  booktitle = {Proceedings of the 36th International Conference on Machine Learning},
  year      = {2019},
  series    = {Proceedings of Machine Learning Research},
  volume    = {97},
  pages     = {6105--6114},
  publisher = {PMLR},
  url       = {https://proceedings.mlr.press/v97/tan19a.html}
}

@ARTICLE{burke_1957,
       author = {{Burke}, Bernard F.},
        title = "{Systematic distortion of the outer regions of the galaxy.}",
      journal = {\aj},
         year = 1957,
        month = may,
       volume = {62},
        pages = {90},
          doi = {10.1086/107463},
       adsurl = {https://ui.adsabs.harvard.edu/abs/1957AJ.....62...90B}
}

@ARTICLE{Djorgovski_1989,
       author = {{Djorgovski}, S. and {Sosin}, Craig},
        title = "{The Warp of the Galactic Stellar Disk Detected in IRAS Source Counts}",
      journal = {\apjl},
         year = 1989,
        month = jun,
       volume = {341},
        pages = {L13},
          doi = {10.1086/185446},
       adsurl = {https://ui.adsabs.harvard.edu/abs/1989ApJ...341L..13D}
}

@ARTICLE{Reshetnikov_1995,
       author = {{Reshetnikov}, V.~P.},
        title = "{On the Statistics of Galactic WARPS}",
      journal = {Astronomical and Astrophysical Transactions},
         year = 1995,
        month = jan,
       volume = {8},
       number = {1},
        pages = {31-37},
          doi = {10.1080/10556799508203294},
       adsurl = {https://ui.adsabs.harvard.edu/abs/1995A&AT....8...31R}
}

@ARTICLE{Reshetnikov_1998,
       author = {{Reshetnikov}, Vladimir and {Combes}, Francoise},
        title = "{Statistics of optical WARPS in spiral disks}",
      journal = {\aap},
         year = 1998,
        month = sep,
       volume = {337},
        pages = {9-16},
          doi = {10.48550/arXiv.astro-ph/9806114},
archivePrefix = {arXiv},
       eprint = {astro-ph/9806114},
 primaryClass = {astro-ph},
       adsurl = {https://ui.adsabs.harvard.edu/abs/1998A&A...337....9R}
}

@ARTICLE{Reshetnikov_1999,
       author = {{Reshetnikov}, Vladimir and {Combes}, Francoise},
        title = "{Statistics of optical WARPS in spiral disks}",
      journal = {\aap},
         year = 1998,
        month = sep,
       volume = {337},
        pages = {9-16},
          doi = {10.48550/arXiv.astro-ph/9806114},
archivePrefix = {arXiv},
       eprint = {astro-ph/9806114},
 primaryClass = {astro-ph},
       adsurl = {https://ui.adsabs.harvard.edu/abs/1998A&A...337....9R}
}

@ARTICLE{Garcia_2002,
       author = {{Garc{\'\i}a-Ruiz}, I. and {Kuijken}, K. and {Dubinski}, J.},
        title = "{The warp of the Galaxy and the Large Magellanic Cloud}",
      journal = {\mnras},
         year = 2002,
        month = dec,
       volume = {337},
       number = {2},
        pages = {459-469},
          doi = {10.1046/j.1365-8711.2002.05923.x},
       adsurl = {https://ui.adsabs.harvard.edu/abs/2002MNRAS.337..459G}
}

@ARTICLE{Saroon_2022,
       author = {{Saroon}, S. and {Subramanian}, S.},
        title = "{Shape of the outer stellar warp in the Large Magellanic Cloud disk}",
      journal = {\aap},
         year = 2022,
        month = oct,
       volume = {666},
          eid = {A103},
        pages = {A103},
          doi = {10.1051/0004-6361/202141435},
archivePrefix = {arXiv},
       eprint = {2207.13269},
 primaryClass = {astro-ph.GA},
       adsurl = {https://ui.adsabs.harvard.edu/abs/2022A&A...666A.103S}
}

@INPROCEEDINGS{Nurhidayat_2020,
       author = {{Nurhidayat}, Rizky Maulana and {Arifyanto}, Mochamad Ikbal and {Puspitarini}, Lucky},
        title = "{Galactic warp from the kinematics of OB stars}",
    booktitle = {European Physical Journal Web of Conferences},
         year = 2020,
       series = {European Physical Journal Web of Conferences},
       volume = {240},
        month = dec,
          eid = {07008},
        pages = {07008},
          doi = {10.1051/epjconf/202024007008},
       adsurl = {https://ui.adsabs.harvard.edu/abs/2020EPJWC.24007008N}
}

@ARTICLE{Hunt_2025,
       author = {{Hunt}, Jason A.~S. and {Vasiliev}, Eugene},
        title = "{Milky Way dynamics in light of Gaia}",
      journal = {\nar},
         year = 2025,
        month = jun,
       volume = {100},
          eid = {101721},
        pages = {101721},
          doi = {10.1016/j.newar.2024.101721},
archivePrefix = {arXiv},
       eprint = {2501.04075},
 primaryClass = {astro-ph.GA},
       adsurl = {https://ui.adsabs.harvard.edu/abs/2025NewAR.10001721H}
}

@ARTICLE{han_2023b,
       author = {{Han}, Jiwon Jesse and {Semenov}, Vadim and {Conroy}, Charlie and {Hernquist}, Lars},
        title = "{Tilted Dark Halos Are Common and Long-lived, and Can Warp Galactic Disks}",
      journal = {\apjl},
         year = 2023,
        month = nov,
       volume = {957},
       number = {2},
          eid = {L24},
        pages = {L24},
          doi = {10.3847/2041-8213/ad0641},
archivePrefix = {arXiv},
       eprint = {2309.07208},
 primaryClass = {astro-ph.GA},
       adsurl = {https://ui.adsabs.harvard.edu/abs/2023ApJ...957L..24H}
}

@ARTICLE{han_2023a,
       author = {{Han}, Jiwon Jesse and {Conroy}, Charlie and {Hernquist}, Lars},
        title = "{A tilted dark halo origin of the Galactic disk warp and flare}",
      journal = {Nature Astronomy},
         year = 2023,
        month = dec,
       volume = {7},
        pages = {1481-1485},
          doi = {10.1038/s41550-023-02076-9},
archivePrefix = {arXiv},
       eprint = {2309.07209},
 primaryClass = {astro-ph.GA},
       adsurl = {https://ui.adsabs.harvard.edu/abs/2023NatAs...7.1481H}
}

@ARTICLE{binney_2024,
       author = {{Binney}, James},
        title = "{Disc distortion revisited}",
      journal = {\mnras},
         year = 2024,
        month = dec,
       volume = {535},
       number = {2},
        pages = {1898-1912},
          doi = {10.1093/mnras/stae2481},
archivePrefix = {arXiv},
       eprint = {2411.04879},
 primaryClass = {astro-ph.GA},
       adsurl = {https://ui.adsabs.harvard.edu/abs/2024MNRAS.535.1898B}
}

@ARTICLE{briggs_1990,
       author = {{Briggs}, F.~H.},
        title = "{Rules of Behavior for Galactic WARPS}",
      journal = {\apj},
         year = 1990,
        month = mar,
       volume = {352},
        pages = {15},
          doi = {10.1086/168512},
       adsurl = {https://ui.adsabs.harvard.edu/abs/1990ApJ...352...15B}
}

@ARTICLE{kerr_1957,
       author = {{Kerr}, F.~J.},
        title = "{A Magellanic effect on the galaxy.}",
      journal = {\aj},
         year = 1957,
        month = may,
       volume = {62},
        pages = {93-93},
          doi = {10.1086/107466},
       adsurl = {https://ui.adsabs.harvard.edu/abs/1957AJ.....62...93K}
}

@ARTICLE{Debattista_1999,
       author = {{Debattista}, Victor P. and {Sellwood}, J.~A.},
        title = "{Warped Galaxies from Misaligned Angular Momenta}",
      journal = {\apjl},
         year = 1999,
        month = mar,
       volume = {513},
       number = {2},
        pages = {L107-L110},
          doi = {10.1086/311913},
archivePrefix = {arXiv},
       eprint = {astro-ph/9901153},
 primaryClass = {astro-ph},
       adsurl = {https://ui.adsabs.harvard.edu/abs/1999ApJ...513L.107D}
}

@ARTICLE{Laporte_2018,
       author = {{Laporte}, Chervin F.~P. and {Johnston}, Kathryn V. and {G{\'o}mez}, Facundo A. and {Garavito-Camargo}, Nicolas and {Besla}, Gurtina},
        title = "{The influence of Sagittarius and the Large Magellanic Cloud on the stellar disc of the Milky Way Galaxy}",
      journal = {\mnras},
         year = 2018,
        month = nov,
       volume = {481},
       number = {1},
        pages = {286-306},
          doi = {10.1093/mnras/sty1574},
archivePrefix = {arXiv},
       eprint = {1710.02538},
 primaryClass = {astro-ph.GA},
       adsurl = {https://ui.adsabs.harvard.edu/abs/2018MNRAS.481..286L}
}

@ARTICLE{Weinberg_1998,
       author = {{Weinberg}, Martin D.},
        title = "{Dynamics of an interacting luminous disc, dark halo and satellite companion}",
      journal = {\mnras},
         year = 1998,
        month = sep,
       volume = {299},
       number = {2},
        pages = {499-514},
          doi = {10.1046/j.1365-8711.1998.01790.x},
       adsurl = {https://ui.adsabs.harvard.edu/abs/1998MNRAS.299..499W}
}

@ARTICLE{Sparke_1998,
       author = {{Sparke}, Linda S. and {Casertano}, Stefano},
        title = "{A model for persistent galactic warps.}",
      journal = {\mnras},
         year = 1988,
        month = oct,
       volume = {234},
        pages = {873-898},
          doi = {10.1093/mnras/234.4.873},
       adsurl = {https://ui.adsabs.harvard.edu/abs/1988MNRAS.234..873S}
}

@ARTICLE{Rok_2010,
       author = {{Ro{\v{s}}kar}, Rok and {Debattista}, Victor P. and {Brooks}, Alyson M. and {Quinn}, Thomas R. and {Brook}, Chris B. and {Governato}, Fabio and {Dalcanton}, Julianne J. and {Wadsley}, James},
        title = "{Misaligned angular momentum in hydrodynamic cosmological simulations: warps, outer discs and thick discs}",
      journal = {\mnras},
         year = 2010,
        month = oct,
       volume = {408},
       number = {2},
        pages = {783-796},
          doi = {10.1111/j.1365-2966.2010.17178.x},
archivePrefix = {arXiv},
       eprint = {1006.1659},
 primaryClass = {astro-ph.CO},
       adsurl = {https://ui.adsabs.harvard.edu/abs/2010MNRAS.408..783R}
}

@ARTICLE{Voort_2015,
       author = {{van de Voort}, Freeke and {Davis}, Timothy A. and {Kere{\v{s}}}, Du{\v{s}}an and {Quataert}, Eliot and {Faucher-Gigu{\`e}re}, Claude-Andr{\'e} and {Hopkins}, Philip F.},
        title = "{The creation and persistence of a misaligned gas disc in a simulated early-type galaxy}",
      journal = {\mnras},
         year = 2015,
        month = aug,
       volume = {451},
       number = {3},
        pages = {3269-3277},
          doi = {10.1093/mnras/stv1217},
archivePrefix = {arXiv},
       eprint = {1504.03685},
 primaryClass = {astro-ph.GA},
       adsurl = {https://ui.adsabs.harvard.edu/abs/2015MNRAS.451.3269V}
}

@ARTICLE{Sellwood_1996,
       author = {{Sellwood}, J.~A.},
        title = "{Axisymmetric Bending Oscillations of Stellar Disks}",
      journal = {\apj},
         year = 1996,
        month = dec,
       volume = {473},
        pages = {733},
          doi = {10.1086/178185},
archivePrefix = {arXiv},
       eprint = {astro-ph/9604123},
 primaryClass = {astro-ph},
       adsurl = {https://ui.adsabs.harvard.edu/abs/1996ApJ...473..733S}
}

@ARTICLE{Binney_1992,
       author = {{Binney}, James},
        title = "{Warps.}",
      journal = {\araa},
         year = 1992,
        month = jan,
       volume = {30},
        pages = {51-74},
          doi = {10.1146/annurev.aa.30.090192.000411},
       adsurl = {https://ui.adsabs.harvard.edu/abs/1992ARA&A..30...51B}
}

@ARTICLE{sheth_2005,
       author = {{Sheth}, Kartik and {Vogel}, Stuart N. and {Regan}, Michael W. and {Thornley}, Michele D. and {Teuben}, Peter J.},
        title = "{Secular Evolution via Bar-driven Gas Inflow: Results from BIMA SONG}",
      journal = {\apj},
         year = 2005,
        month = oct,
       volume = {632},
       number = {1},
        pages = {217-226},
          doi = {10.1086/432409},
archivePrefix = {arXiv},
       eprint = {astro-ph/0505393},
 primaryClass = {astro-ph},
       adsurl = {https://ui.adsabs.harvard.edu/abs/2005ApJ...632..217S}
}

@ARTICLE{sakamoto_1999,
       author = {{Sakamoto}, K. and {Okumura}, S.~K. and {Ishizuki}, S. and {Scoville}, N.~Z.},
        title = "{Bar-driven Transport of Molecular Gas to Galactic Centers and Its Consequences}",
      journal = {\apj},
         year = 1999,
        month = nov,
       volume = {525},
       number = {2},
        pages = {691-701},
          doi = {10.1086/307910},
archivePrefix = {arXiv},
       eprint = {astro-ph/9906454},
 primaryClass = {astro-ph},
       adsurl = {https://ui.adsabs.harvard.edu/abs/1999ApJ...525..691S}
}

@ARTICLE{shlosman_1989,
       author = {{Shlosman}, Isaac and {Frank}, Juhan and {Begelman}, Mitchell C.},
        title = "{Bars within bars: a mechanism for fuelling active galactic nuclei}",
      journal = {\nat},
         year = 1989,
        month = mar,
       volume = {338},
       number = {6210},
        pages = {45-47},
          doi = {10.1038/338045a0},
       adsurl = {https://ui.adsabs.harvard.edu/abs/1989Natur.338...45S}
}

@ARTICLE{kormendy_2004,
       author = {{Kormendy}, John and {Kennicutt}, Jr., Robert C.},
        title = "{Secular Evolution and the Formation of Pseudobulges in Disk Galaxies}",
      journal = {\araa},
         year = 2004,
        month = sep,
       volume = {42},
       number = {1},
        pages = {603-683},
          doi = {10.1146/annurev.astro.42.053102.134024},
archivePrefix = {arXiv},
       eprint = {astro-ph/0407343},
 primaryClass = {astro-ph},
       adsurl = {https://ui.adsabs.harvard.edu/abs/2004ARA&A..42..603K}
}

@ARTICLE{Bluck_2014,
       author = {{Bluck}, Asa F.~L. and {Mendel}, J. Trevor and {Ellison}, Sara L. and {Moreno}, Jorge and {Simard}, Luc and {Patton}, David R. and {Starkenburg}, Else},
        title = "{Bulge mass is king: the dominant role of the bulge in determining the fraction of passive galaxies in the Sloan Digital Sky Survey}",
      journal = {\mnras},
         year = 2014,
        month = jun,
       volume = {441},
       number = {1},
        pages = {599-629},
          doi = {10.1093/mnras/stu594},
archivePrefix = {arXiv},
       eprint = {1403.5269},
 primaryClass = {astro-ph.GA},
       adsurl = {https://ui.adsabs.harvard.edu/abs/2014MNRAS.441..599B}
}

@ARTICLE{Kormendy_2013,
       author = {{Kormendy}, John and {Ho}, Luis C.},
        title = "{Coevolution (Or Not) of Supermassive Black Holes and Host Galaxies}",
      journal = {\araa},
         year = 2013,
        month = aug,
       volume = {51},
       number = {1},
        pages = {511-653},
          doi = {10.1146/annurev-astro-082708-101811},
archivePrefix = {arXiv},
       eprint = {1304.7762},
 primaryClass = {astro-ph.CO},
       adsurl = {https://ui.adsabs.harvard.edu/abs/2013ARA&A..51..511K}
}

@ARTICLE{jonsson_2024,
       author = {{J{\'o}nsson}, Viktor Hrannar and {McMillan}, Paul J.},
        title = "{The tangled warp of the Milky Way}",
      journal = {\aap},
         year = 2024,
        month = aug,
       volume = {688},
          eid = {A38},
        pages = {A38},
          doi = {10.1051/0004-6361/202449744},
archivePrefix = {arXiv},
       eprint = {2405.09624},
 primaryClass = {astro-ph.GA},
       adsurl = {https://ui.adsabs.harvard.edu/abs/2024A&A...688A..38J}
}

@ARTICLE{volker_1999,
       author = {{Springel}, Volker and {White}, Simon D.~M.},
        title = "{Tidal tails in cold dark matter cosmologies}",
      journal = {\mnras},
         year = 1999,
        month = jul,
       volume = {307},
       number = {1},
        pages = {162-178},
          doi = {10.1046/j.1365-8711.1999.02613.x},
archivePrefix = {arXiv},
       eprint = {astro-ph/9807320},
 primaryClass = {astro-ph},
       adsurl = {https://ui.adsabs.harvard.edu/abs/1999MNRAS.307..162S}
}

@ARTICLE{Ibata_2001,
       author = {{Ibata}, Rodrigo and {Lewis}, Geraint F. and {Irwin}, Michael and {Totten}, Edward and {Quinn}, Thomas},
        title = "{Great Circle Tidal Streams: Evidence for a Nearly Spherical Massive Dark Halo around the Milky Way}",
      journal = {\apj},
         year = 2001,
        month = apr,
       volume = {551},
       number = {1},
        pages = {294-311},
          doi = {10.1086/320060},
archivePrefix = {arXiv},
       eprint = {astro-ph/0004011},
 primaryClass = {astro-ph},
       adsurl = {https://ui.adsabs.harvard.edu/abs/2001ApJ...551..294I}
}

@ARTICLE{johnston_1999,
       author = {{Johnston}, Kathryn V. and {Zhao}, HongSheng and {Spergel}, David N. and {Hernquist}, Lars},
        title = "{Tidal Streams as Probes of the Galactic Potential}",
      journal = {\apjl},
         year = 1999,
        month = feb,
       volume = {512},
       number = {2},
        pages = {L109-L112},
          doi = {10.1086/311876},
archivePrefix = {arXiv},
       eprint = {astro-ph/9807243},
 primaryClass = {astro-ph},
       adsurl = {https://ui.adsabs.harvard.edu/abs/1999ApJ...512L.109J}
}

@ARTICLE{speagle_2014,
       author = {{Speagle}, J.~S. and {Steinhardt}, C.~L. and {Capak}, P.~L. and {Silverman}, J.~D.},
        title = "{A Highly Consistent Framework for the Evolution of the Star-Forming ``Main Sequence'' from z \raisebox{-0.5ex}\textasciitilde 0-6}",
      journal = {\apjs},
         year = 2014,
        month = oct,
       volume = {214},
       number = {2},
          eid = {15},
        pages = {15},
          doi = {10.1088/0067-0049/214/2/15},
archivePrefix = {arXiv},
       eprint = {1405.2041},
 primaryClass = {astro-ph.GA},
       adsurl = {https://ui.adsabs.harvard.edu/abs/2014ApJS..214...15S}
}

@ARTICLE{optuna_2019,
       author = {{Akiba}, Takuya and {Sano}, Shotaro and {Yanase}, Toshihiko and {Ohta}, Takeru and {Koyama}, Masanori},
        title = "{Optuna: A Next-generation Hyperparameter Optimization Framework}",
      journal = {arXiv e-prints},
         year = 2019,
        month = jul,
          eid = {arXiv:1907.10902},
        pages = {arXiv:1907.10902},
          doi = {10.48550/arXiv.1907.10902},
archivePrefix = {arXiv},
       eprint = {1907.10902},
 primaryClass = {cs.LG},
       adsurl = {https://ui.adsabs.harvard.edu/abs/2019arXiv190710902A}
}

@article{Sanchez_2003,
  author = {Sánchez-Saavedra, M. L. and Battaner, E. and Guijarro, A.},
  title = {A catalog of optical warps in spiral and lenticular galaxies},
  journal = {Astronomy \& Astrophysics},
  volume = {399},
  pages = {457--469},
  year = {2003}
}

@article{Ann_2006,
  author = {Ann, H. B. and Park, J.-C.},
  title = {Warped disks in spiral galaxies},
  journal = {New Astronomy},
  volume = {11},
  pages = {293--303},
  year = {2006}
}

@article{Skryabina_2024,
  author = {Skryabina, A. and Reshetnikov, V. and Sotnikova, N. and others},
  title = {Structural asymmetries and warps in edge-on galaxies from deep imaging surveys},
  journal = {Monthly Notices of the Royal Astronomical Society},
  year = {2024},
  note = {in press}
}

@ARTICLE{Abazajian_2009,
       author = {{Abazajian}, Kevork N. and {Adelman-McCarthy}, Jennifer K. and {Ag{\"u}eros}, Marcel A. and {Allam}, Sahar S. and {Allende Prieto}, Carlos and {An}, Deokkeun and {Anderson}, Kurt S.~J. and {Anderson}, Scott F. and {Annis}, James and {Bahcall}, Neta A. and {Bailer-Jones}, C.~A.~L. and {Barentine}, J.~C. and {Bassett}, Bruce A. and {Becker}, Andrew C. and {Beers}, Timothy C. and {Bell}, Eric F. and {Belokurov}, Vasily and {Berlind}, Andreas A. and {Berman}, Eileen F. and {Bernardi}, Mariangela and {Bickerton}, Steven J. and {Bizyaev}, Dmitry and {Blakeslee}, John P. and {Blanton}, Michael R. and {Bochanski}, John J. and {Boroski}, William N. and {Brewington}, Howard J. and {Brinchmann}, Jarle and {Brinkmann}, J. and {Brunner}, Robert J. and {Budav{\'a}ri}, Tam{\'a}s and {Carey}, Larry N. and {Carliles}, Samuel and {Carr}, Michael A. and {Castander}, Francisco J. and {Cinabro}, David and {Connolly}, A.~J. and {Csabai}, Istv{\'a}n and {Cunha}, Carlos E. and {Czarapata}, Paul C. and {Davenport}, James R.~A. and {de Haas}, Ernst and {Dilday}, Ben and {Doi}, Mamoru and {Eisenstein}, Daniel J. and {Evans}, Michael L. and {Evans}, N.~W. and {Fan}, Xiaohui and {Friedman}, Scott D. and {Frieman}, Joshua A. and {Fukugita}, Masataka and {G{\"a}nsicke}, Boris T. and {Gates}, Evalyn and {Gillespie}, Bruce and {Gilmore}, G. and {Gonzalez}, Belinda and {Gonzalez}, Carlos F. and {Grebel}, Eva K. and {Gunn}, James E. and {Gy{\"o}ry}, Zsuzsanna and {Hall}, Patrick B. and {Harding}, Paul and {Harris}, Frederick H. and {Harvanek}, Michael and {Hawley}, Suzanne L. and {Hayes}, Jeffrey J.~E. and {Heckman}, Timothy M. and {Hendry}, John S. and {Hennessy}, Gregory S. and {Hindsley}, Robert B. and {Hoblitt}, J. and {Hogan}, Craig J. and {Hogg}, David W. and {Holtzman}, Jon A. and {Hyde}, Joseph B. and {Ichikawa}, Shin-ichi and {Ichikawa}, Takashi and {Im}, Myungshin and {Ivezi{\'c}}, {\v{Z}}eljko and {Jester}, Sebastian and {Jiang}, Linhua and {Johnson}, Jennifer A. and {Jorgensen}, Anders M. and {Juri{\'c}}, Mario and {Kent}, Stephen M. and {Kessler}, R. and {Kleinman}, S.~J. and {Knapp}, G.~R. and {Konishi}, Kohki and {Kron}, Richard G. and {Krzesinski}, Jurek and {Kuropatkin}, Nikolay and {Lampeitl}, Hubert and {Lebedeva}, Svetlana and {Lee}, Myung Gyoon and {Lee}, Young Sun and {French Leger}, R. and {L{\'e}pine}, S{\'e}bastien and {Li}, Nolan and {Lima}, Marcos and {Lin}, Huan and {Long}, Daniel C. and {Loomis}, Craig P. and {Loveday}, Jon and {Lupton}, Robert H. and {Magnier}, Eugene and {Malanushenko}, Olena and {Malanushenko}, Viktor and {Mandelbaum}, Rachel and {Margon}, Bruce and {Marriner}, John P. and {Mart{\'\i}nez-Delgado}, David and {Matsubara}, Takahiko and {McGehee}, Peregrine M. and {McKay}, Timothy A. and {Meiksin}, Avery and {Morrison}, Heather L. and {Mullally}, Fergal and {Munn}, Jeffrey A. and {Murphy}, Tara and {Nash}, Thomas and {Nebot}, Ada and {Neilsen}, Jr., Eric H. and {Newberg}, Heidi Jo and {Newman}, Peter R. and {Nichol}, Robert C. and {Nicinski}, Tom and {Nieto-Santisteban}, Maria and {Nitta}, Atsuko and {Okamura}, Sadanori and {Oravetz}, Daniel J. and {Ostriker}, Jeremiah P. and {Owen}, Russell and {Padmanabhan}, Nikhil and {Pan}, Kaike and {Park}, Changbom and {Pauls}, George and {Peoples}, Jr., John and {Percival}, Will J. and {Pier}, Jeffrey R. and {Pope}, Adrian C. and {Pourbaix}, Dimitri and {Price}, Paul A. and {Purger}, Norbert and {Quinn}, Thomas and {Raddick}, M. Jordan and {Re Fiorentin}, Paola and {Richards}, Gordon T. and {Richmond}, Michael W. and {Riess}, Adam G. and {Rix}, Hans-Walter and {Rockosi}, Constance M. and {Sako}, Masao and {Schlegel}, David J. and {Schneider}, Donald P. and {Scholz}, Ralf-Dieter and {Schreiber}, Matthias R. and {Schwope}, Axel D. and {Seljak}, Uro{\v{s}} and {Sesar}, Branimir and {Sheldon}, Erin and {Shimasaku}, Kazu and {Sibley}, Valena C. and {Simmons}, A.~E. and {Sivarani}, Thirupathi and {Allyn Smith}, J. and {Smith}, Martin C. and {Smol{\v{c}}i{\'c}}, Vernesa and {Snedden}, Stephanie A. and {Stebbins}, Albert and {Steinmetz}, Matthias and {Stoughton}, Chris and {Strauss}, Michael A. and {SubbaRao}, Mark and {Suto}, Yasushi and {Szalay}, Alexander S. and {Szapudi}, Istv{\'a}n and {Szkody}, Paula and {Tanaka}, Masayuki and {Tegmark}, Max and {Teodoro}, Luis F.~A. and {Thakar}, Aniruddha R. and {Tremonti}, Christy A. and {Tucker}, Douglas L. and {Uomoto}, Alan and {Vanden Berk}, Daniel E. and {Vandenberg}, Jan and {Vidrih}, S. and {Vogeley}, Michael S. and {Voges}, Wolfgang and {Vogt}, Nicole P. and {Wadadekar}, Yogesh and {Watters}, Shannon and {Weinberg}, David H. and {West}, Andrew A. and {White}, Simon D.~M. and {Wilhite}, Brian C. and {Wonders}, Alainna C. and {Yanny}, Brian and {Yocum}, D.~R.},
        title = "{The Seventh Data Release of the Sloan Digital Sky Survey}",
      journal = {\apjs},
         year = 2009,
        month = jun,
       volume = {182},
       number = {2},
        pages = {543-558},
          doi = {10.1088/0067-0049/182/2/543},
archivePrefix = {arXiv},
       eprint = {0812.0649},
 primaryClass = {astro-ph},
       adsurl = {https://ui.adsabs.harvard.edu/abs/2009ApJS..182..543A}
}

@ARTICLE{Aguado_2019,
       author = {{Aguado}, D.~S. and {Ahumada}, Romina and {Almeida}, Andr{\'e}s and {Anderson}, Scott F. and {Andrews}, Brett H. and {Anguiano}, Borja and {Aquino Ort{\'\i}z}, Erik and {Arag{\'o}n-Salamanca}, Alfonso and {Argudo-Fern{\'a}ndez}, Maria and {Aubert}, Marie and {Avila-Reese}, Vladimir and {Badenes}, Carles and {Barboza Rembold}, Sandro and {Barger}, Kat and {Barrera-Ballesteros}, Jorge and {Bates}, Dominic and {Bautista}, Julian and {Beaton}, Rachael L. and {Beers}, Timothy C. and {Belfiore}, Francesco and {Bernardi}, Mariangela and {Bershady}, Matthew and {Beutler}, Florian and {Bird}, Jonathan and {Bizyaev}, Dmitry and {Blanc}, Guillermo A. and {Blanton}, Michael R. and {Blomqvist}, Michael and {Bolton}, Adam S. and {Boquien}, M{\'e}d{\'e}ric and {Borissova}, Jura and {Bovy}, Jo and {Brandt}, William Nielsen and {Brinkmann}, Jonathan and {Brownstein}, Joel R. and {Bundy}, Kevin and {Burgasser}, Adam and {Byler}, Nell and {Cano Diaz}, Mariana and {Cappellari}, Michele and {Carrera}, Ricardo and {Cervantes Sodi}, Bernardo and {Chen}, Yanping and {Cherinka}, Brian and {Choi}, Peter Doohyun and {Chung}, Haeun and {Coffey}, Damien and {Comerford}, Julia M. and {Comparat}, Johan and {Covey}, Kevin and {da Silva Ilha}, Gabriele and {da Costa}, Luiz and {Dai}, Yu Sophia and {Damke}, Guillermo and {Darling}, Jeremy and {Davies}, Roger and {Dawson}, Kyle and {de Sainte Agathe}, Victoria and {Deconto Machado}, Alice and {Del Moro}, Agnese and {De Lee}, Nathan and {Diamond-Stanic}, Aleksandar M. and {Dom{\'\i}nguez S{\'a}nchez}, Helena and {Donor}, John and {Drory}, Niv and {du Mas des Bourboux}, H{\'e}lion and {Duckworth}, Chris and {Dwelly}, Tom and {Ebelke}, Garrett and {Emsellem}, Eric and {Escoffier}, Stephanie and {Fern{\'a}ndez-Trincado}, Jos{\'e} G. and {Feuillet}, Diane and {Fischer}, Johanna-Laina and {Fleming}, Scott W. and {Fraser-McKelvie}, Amelia and {Freischlad}, Gordon and {Frinchaboy}, Peter M. and {Fu}, Hai and {Galbany}, Llu{\'\i}s and {Garcia-Dias}, Rafael and {Garc{\'\i}a-Hern{\'a}ndez}, D.~A. and {Garma Oehmichen}, Luis Alberto and {Geimba Maia}, Marcio Antonio and {Gil-Mar{\'\i}n}, H{\'e}ctor and {Grabowski}, Kathleen and {Gu}, Meng and {Guo}, Hong and {Ha}, Jaewon and {Harrington}, Emily and {Hasselquist}, Sten and {Hayes}, Christian R. and {Hearty}, Fred and {Hernandez Toledo}, Hector and {Hicks}, Harry and {Hogg}, David W. and {Holley-Bockelmann}, Kelly and {Holtzman}, Jon A. and {Hsieh}, Bau-Ching and {Hunt}, Jason A.~S. and {Hwang}, Ho Seong and {Ibarra-Medel}, H{\'e}ctor J. and {Jimenez Angel}, Camilo Eduardo and {Johnson}, Jennifer and {Jones}, Amy and {J{\"o}nsson}, Henrik and {Kinemuchi}, Karen and {Kollmeier}, Juna and {Krawczyk}, Coleman and {Kreckel}, Kathryn and {Kruk}, Sandor and {Lacerna}, Ivan and {Lan}, Ting-Wen and {Lane}, Richard R. and {Law}, David R. and {Lee}, Young-Bae and {Li}, Cheng and {Lian}, Jianhui and {Lin}, Lihwai and {Lin}, Yen-Ting and {Lintott}, Chris and {Long}, Dan and {Longa-Pe{\~n}a}, Pen{\'e}lope and {Mackereth}, J. Ted and {de la Macorra}, Axel and {Majewski}, Steven R. and {Malanushenko}, Olena and {Manchado}, Arturo and {Maraston}, Claudia and {Mariappan}, Vivek and {Marinelli}, Mariarosa and {Marques-Chaves}, Rui and {Masseron}, Thomas and {Masters}, Karen L. and {McDermid}, Richard M. and {Medina Pe{\~n}a}, Nicol{\'a}s and {Meneses-Goytia}, Sofia and {Merloni}, Andrea and {Merrifield}, Michael and {Meszaros}, Szabolcs and {Minniti}, Dante and {Minsley}, Rebecca and {Muna}, Demitri and {Myers}, Adam D. and {Nair}, Preethi and {Correa do Nascimento}, Janaina and {Newman}, Jeffrey A. and {Nitschelm}, Christian and {Olmstead}, Matthew D. and {Oravetz}, Audrey and {Oravetz}, Daniel and {Ortega Minakata}, Ren{\'e} A. and {Pace}, Zach and {Padilla}, Nelson and {Palicio}, Pedro A. and {Pan}, Kaike and {Pan}, Hsi-An and {Parikh}, Taniya and {Parker}, III, James and {Peirani}, Sebastien and {Penny}, Samantha and {Percival}, Will J. and {Perez-Fournon}, Ismael and {Peterken}, Thomas and {Pinsonneault}, Marc H. and {Prakash}, Abhishek and {Raddick}, M. Jordan and {Raichoor}, Anand and {Riffel}, Rogemar A. and {Riffel}, Rog{\'e}rio and {Rix}, Hans-Walter and {Robin}, Annie C. and {Roman-Lopes}, Alexandre and {Rose}, Benjamin and {Ross}, Ashley J. and {Rossi}, Graziano and {Rowlands}, Kate and {Rubin}, Kate H.~R. and {S{\'a}nchez}, Sebasti{\'a}n F. and {S{\'a}nchez-Gallego}, Jos{\'e} R. and {Sayres}, Conor and {Schaefer}, Adam and {Schiavon}, Ricardo P. and {Schimoia}, Jaderson S. and {Schlafly}, Edward and {Schlegel}, David and {Schneider}, Donald P. and {Schultheis}, Mathias and {Seo}, Hee-Jong and {Shamsi}, Shoaib J. and {Shao}, Zhengyi and {Shen}, Shiyin and {Shetty}, Shravan and {Simonian}, Gregory and {Smethurst}, Rebecca J. and {Sobeck}, Jennifer and {Souter}, Barbara J. and {Spindler}, Ashley and {Stark}, David V. and {Stassun}, Keivan G.},
        title = "{The Fifteenth Data Release of the Sloan Digital Sky Surveys: First Release of MaNGA-derived Quantities, Data Visualization Tools, and Stellar Library}",
      journal = {\apjs},
         year = 2019,
        month = feb,
       volume = {240},
       number = {2},
          eid = {23},
        pages = {23},
          doi = {10.3847/1538-4365/aaf651},
archivePrefix = {arXiv},
       eprint = {1812.02759},
 primaryClass = {astro-ph.IM},
       adsurl = {https://ui.adsabs.harvard.edu/abs/2019ApJS..240...23A}
}

@ARTICLE{zhou_2021,
       author = {{Zhou}, Rongpu and {Newman}, Jeffrey A. and {Mao}, Yao-Yuan and {Meisner}, Aaron and {Moustakas}, John and {Myers}, Adam D. and {Prakash}, Abhishek and {Zentner}, Andrew R. and {Brooks}, David and {Duan}, Yutong and {Landriau}, Martin and {Levi}, Michael E. and {Prada}, Francisco and {Tarle}, Gregory},
        title = "{The clustering of DESI-like luminous red galaxies using photometric redshifts}",
      journal = {\mnras},
         year = 2021,
        month = mar,
       volume = {501},
       number = {3},
        pages = {3309-3331},
          doi = {10.1093/mnras/staa3764},
archivePrefix = {arXiv},
       eprint = {2001.06018},
 primaryClass = {astro-ph.CO},
       adsurl = {https://ui.adsabs.harvard.edu/abs/2021MNRAS.501.3309Z}
}

@ARTICLE{haynes_2018,
       author = {{Haynes}, Martha P. and {Giovanelli}, Riccardo and {Kent}, Brian R. and {Adams}, Elizabeth A.~K. and {Balonek}, Thomas J. and {Craig}, David W. and {Fertig}, Derek and {Finn}, Rose and {Giovanardi}, Carlo and {Hallenbeck}, Gregory and {Hess}, Kelley M. and {Hoffman}, G. Lyle and {Huang}, Shan and {Jones}, Michael G. and {Koopmann}, Rebecca A. and {Kornreich}, David A. and {Leisman}, Lukas and {Miller}, Jeffrey and {Moorman}, Crystal and {O'Connor}, Jessica and {O'Donoghue}, Aileen and {Papastergis}, Emmanouil and {Troischt}, Parker and {Stark}, David and {Xiao}, Li},
        title = "{The Arecibo Legacy Fast ALFA Survey: The ALFALFA Extragalactic H I Source Catalog}",
      journal = {\apj},
         year = 2018,
        month = jul,
       volume = {861},
       number = {1},
          eid = {49},
        pages = {49},
          doi = {10.3847/1538-4357/aac956},
archivePrefix = {arXiv},
       eprint = {1805.11499},
 primaryClass = {astro-ph.GA},
       adsurl = {https://ui.adsabs.harvard.edu/abs/2018ApJ...861...49H}
}

@ARTICLE{yang_2007,
       author = {{Yang}, Xiaohu and {Mo}, H.~J. and {van den Bosch}, Frank C. and {Pasquali}, Anna and {Li}, Cheng and {Barden}, Marco},
        title = "{Galaxy Groups in the SDSS DR4. I. The Catalog and Basic Properties}",
      journal = {\apj},
         year = 2007,
        month = dec,
       volume = {671},
       number = {1},
        pages = {153-170},
          doi = {10.1086/522027},
archivePrefix = {arXiv},
       eprint = {0707.4640},
 primaryClass = {astro-ph},
       adsurl = {https://ui.adsabs.harvard.edu/abs/2007ApJ...671..153Y}
}
\bibliographystyle{aasjournal}

\end{document}